\documentclass[11pt,a4paper]{article}
\usepackage[T1]{fontenc}
\usepackage[utf8]{inputenc}
\usepackage{lmodern}
\usepackage{amsmath,amssymb,booktabs,needspace}
\usepackage[margin=0.8in]{geometry}
\usepackage{microtype}
\usepackage{graphicx,array,float}
\usepackage{hyperref}
\usepackage{bookmark}
\usepackage{orcidlink}
\newcommand{\email}[1]{\thanks{\href{mailto:#1}{\texttt{#1}}}}
\hypersetup{hidelinks,pdfauthor={Song He},pdftitle={Simplicity of the three-point stress-tensor form factor beyond the planar limit}}
\allowdisplaybreaks[2]
\numberwithin{equation}{section}
\title{\Large\bfseries Simplicity of the three-point stress-tensor form factor\\
beyond the planar limit}
\author{Song He\,\orcidlink{0000-0002-9534-3844}\email{songhe@itp.ac.cn}\\[0.5em]
\small New Cornerstone Laboratory, Institute of Theoretical Physics,\\[-0.1em]
\small Chinese Academy of Sciences, Beijing 100190, China}
\date{}
\begin{document}
\maketitle
\vspace{-1.3em}
\begin{abstract}
We express the complete three-loop subleading-color finite remainder of the
three-point stress-tensor-multiplet form factor in $\mathcal N=4$ super
Yang--Mills theory as a permutation sum of a single $A_2$-type function. The
non-product part of the seed is a combination of 35 weight-six generalized
polylogarithms, while all lower-weight products and zeta-value terms belong
to the same five-letter algebra. A bootstrap based on independently computed
collinear data, branch-cut constraints and empirical final-entry relations
determines both the symbol and the full function. The remaining
function-level ambiguity is proportional to $\zeta_2$ times the two-loop
planar remainder and is fixed by a single final-entry relation. The
weight-three left coproduct factors of the planar and subleading-color
results span the same space modulo products; at weight four, the
subleading-color result contains four additional nonclassical directions.
Through the established maximal-transcendentality correspondence, this
compact representation also describes the weight-six subleading-color sector
of Higgs-plus-three-gluon amplitudes in QCD. We also construct a four-loop
candidate symbol space under a conjectural extension of the three-loop
constraints.
\end{abstract}

\section{Introduction}
\label{sec:intro}
Multi-loop amplitudes and form factors are often much simpler than the
individual Feynman integrals that contribute to them~\cite{GSVV,Form2,DGR}.
Cancellations in the sum can reduce the symbol alphabet and leave a
compact polylogarithmic expression. In planar $\mathcal N=4$ super
Yang--Mills theory, constraints on symbol alphabets and entries,
together with physical boundary data, have enabled bootstrap
calculations at high loop order~\cite{DMW,EightLoops}.
Subleading-color corrections test which of these structures depend on
planar color ordering and which survive in the full theory. The
full-color two-loop five-point amplitude~\cite{FullColor5} provides one
such test. Here we study the three-loop three-point stress-tensor
form factor.

The three-point stress-tensor-multiplet form factor first receives a
subleading-color (SLC) correction at three loops. Dividing out the
tree-level factors and subtracting the infrared poles gives, at fixed
overall scale, a weight-six finite remainder depending on two independent
kinematic ratios. Guan, Mistlberger and Ruf computed this remainder using
integration-by-parts reduction and differential equations~\cite{GMR}.
Although the individual integrals involve a larger alphabet, the answer
uses only six rational letters and has small integer symbol
coefficients. The planar adjacency restriction is violated only in the final pair
of symbol entries. We use these properties to reconstruct and simplify the complete
finite remainder.

Our main result is a permutation-sum representation of the complete
finite remainder in terms of a single $A_2$-type seed. Its non-product
part consists of 35 weight-six generalized polylogarithms (GPLs), given
in Section~\ref{sec:uplift}. All lower-weight products and zeta-value
terms use the same five-letter alphabet and are listed in
Appendix~\ref{app:merged}. The $A_2$ alphabet~\cite{GPSV,PSSV} therefore
organizes the complete seed, including the terms invisible to its
symbol.

The reconstruction uses branch regularity, permutation symmetry,
independently computed collinear data, and four families of
triple-final-entry relations inferred from Ref.~\cite{GMR}. These
conditions uniquely fix the symbol. At the function level, collinear and
soft limits, monodromy constraints and the single-final-entry equation
leave a one-dimensional ambiguity proportional to
$\zeta_2R_{\rm LC}^{(2)}$. A triple-final-entry relation fixes its
coefficient, linking the nonplanar correction to the two-loop planar
remainder.

The compact form does not reduce the SLC correction to a multiple of
the planar remainder plus lower-weight products. At weight three their
coproduct spaces coincide modulo products, whereas at weight four the
SLC result contains four additional nonclassical directions. The
weight-four coproduct components responsible for violations of planar
last-two adjacency are nevertheless classical polylogarithms. Their logarithmic and one-mass-box coproducts give
explicit sources for the departure from planar final-entry conditions.

The companion paper~\cite{FF4SLC} treats the two-loop four-point
form factor at finite color using numerical unitarity, a master-integral
bootstrap and a direct form-factor bootstrap. The present analysis
focuses on the complete three-loop three-point function and the
structure of its final entries.

The result also applies to Higgs amplitudes in QCD. The
maximal-transcendentality relation between form factors and Higgs
amplitudes was found at two loops~\cite{Form2,Higgs2,HiggsSYM} and
verified at three loops, including the SLC terms~\cite{Higgs3}. Let
$h^{(3),Hggg}_{\lambda,\mathrm{SLC}}$ denote the finite scalar coefficient after
factoring out the tree-level helicity factor and the effective Higgs
Wilson coefficient. With coupling, color, scale and subtraction conventions
matched to \eqref{eq:norm},
\begin{equation}
 \mathsf W_6 h^{(3),Hggg}_{+++,\mathrm{SLC}}
 =\mathsf W_6 h^{(3),Hggg}_{-++,\mathrm{SLC}}=48F.
 \label{eq:Higgs3}
\end{equation}
The projection $\mathsf W_6$ includes zeta values multiplying lower-weight
functions. The simplicity found here thus extends beyond $\mathcal N=4$
SYM: by Eq.~\eqref{eq:Higgs3}, the full compact expression, including
its period terms, also describes the highest-weight SLC part of the
QCD amplitude.

After fixing conventions below, we describe the symbol bootstrap in
Section~\ref{sec:symbol} and give the full function in
Section~\ref{sec:uplift}. Section~\ref{sec:comparison} compares the color
sectors and their final-entry structure. Section~\ref{sec:fourloops}
constructs a conjectural four-loop symbol space and studies its
collinear limits. We discuss further directions in
Section~\ref{sec:conclusions}.

\subsection{Three-point form factors and conventions}
\label{sec:setup}
We consider the MHV form factor of the protected scalar operator
$\operatorname{tr}(\phi^2)$ in the chiral stress-tensor
multiplet~\cite{Form1,Form2}, referred to below as the stress-tensor form
factor. For three massless external momenta, with
$q=p_1+p_2+p_3$ and $s_{ij}=(p_i+p_j)^2$, we set
\begin{equation}
 u=\frac{s_{12}}{q^2},\qquad v=\frac{s_{23}}{q^2},\qquad
 w=\frac{s_{31}}{q^2}=1-u-v,\qquad \bar u=1-u,
 \label{eq:kinematics}
\end{equation}
with analogous definitions for $\bar v,\bar w$, and begin in the
pseudo-Euclidean region $q^2,s_{ij}<0$, where $u,v,w>0$.

After extracting the tree-level factors, we write
$\mathcal I(\epsilon)=\mathcal F/\mathcal F^{(0)}
=Z_{\rm IR}(\epsilon)H(\epsilon)$ in $d=4-2\epsilon$, and define
$H=\lim_{\epsilon\to0}Z_{\rm IR}^{-1}\mathcal I$.
We subtract only infrared poles, so that $\log Z_{\rm IR}$ has no
finite term~\cite{IRfactor}; we follow the coupling and loop-measure
conventions of Ref.~\cite{GMR}. The protected operator requires no
ultraviolet renormalization. Thus $H$ is an amplitude-level finite
remainder with pole-only subtraction and tree-level normalization.

The SLC sector first contributes at three loops~\cite{GMR}. With
$a_s=\alpha_sN_c/(4\pi)$, we write the color expansion as
\begin{equation}
 H=H_{\rm LC}+\frac{a_s^3}{N_c^2}\,48F(u,v,w;Q^2)
       +O(a_s^4/N_c^2),
 \qquad Q^2=\frac{-q^2-i0}{\mu^2}.
 \label{eq:hard-color}
\end{equation}
We quote the result at the Euclidean reference scale $Q^2=1$:
\begin{equation}
 F(u,v,w)=\frac{\mathcal I^{(3)}_{\mathrm{SLC,fin}}(u,v,w;Q^2=1)}{48}.
 \label{eq:norm}
\end{equation}
Here $\mathcal I^{(3)}_{\mathrm{SLC,fin}}$ is the coefficient of $a_s^3$
in \texttt{FF2\_SLC\_Finite} supplied with Ref.~\cite{GMR}; its variable
\texttt{Q2} equals $-q^2$ when $\mu^2=1$.
At three loops, the SLC pole arises entirely from the non-dipole
soft anomalous dimension~\cite{ADG,HighPrecision}. The electronic
expressions accompanying Ref.~\cite{GMR} give
\begin{equation}
 \frac{\mathcal I^{(3)}_{\rm SLC}}{48}
 =(Q^2)^{-3\epsilon}
 \left[-\frac{c^{(3)}}{2\epsilon}+F(u,v,w)+O(\epsilon)\right],
 \qquad c^{(3)}=\zeta_5+2\zeta_2\zeta_3.
 \label{eq:slc-pole}
\end{equation}
In the convention of Ref.~\cite{HighPrecision},
$\mathcal I_{\rm SLC}^{(3)}=12\mathcal I_{\rm NP}^{(3)}$, since the
NP contribution carries a factor $12/N_c^2$. Expanding the scale
factor in~\eqref{eq:slc-pole} before pole-only subtraction retains the
finite logarithm,
\begin{equation}
 F(u,v,w;Q^2)=F(u,v,w)+C_5\log Q^2,
 \qquad C_5=\frac32c^{(3)}=3\zeta_2\zeta_3+\frac32\zeta_5.
 \label{eq:scale}
\end{equation}
The finite logarithm in~\eqref{eq:scale} is retained throughout.

For comparison we also use the planar BDS-like quantity $\mathcal E$
and remainder $R$, related by~\cite{BDS,Form2,DMW,EightLoops}
\begin{equation}
 \mathcal E=\exp\!\left[\frac{\Gamma_{\rm cusp}}4\mathcal E^{(1)}+R\right],
 \qquad \mathcal E^{(1)}=2\sum_{i=u,v,w}\operatorname{Li}_2(1-1/i).
 \label{eq:planar-normalizations}
\end{equation}
The finite conversions from $H$ to $R$ and $\mathcal E$ involve
lower-loop products and zeta values. These conversions change the
adjacency and final-entry properties, so we distinguish the two planar
normalizations throughout.

The common six-letter alphabet is
\begin{equation}
 \mathcal A=(u,v,w,\bar u,\bar v,\bar w).
 \label{eq:alphabet}
\end{equation}
Planar form-factor spaces impose physical first entries, integrability,
branch regularity and restrictions on adjacent complementary letters
such as $\bar u\otimes\bar v$~\cite{DMW,EightLoops}. The SLC result retains
the six letters but relaxes adjacency in the last two entries and weakens
the single-final-entry condition~\cite{GMR}. We state the precise restrictions
used in our bootstrap in the next section. All symbol-support and
coefficient comparisons use the ordered basis~\eqref{eq:alphabet}.

For a pure function, our differential convention is
\begin{equation}
 dF=\sum_{a\in\mathcal A}F^a\,d\log a,\qquad F^{a,b}=(F^b)^a.
 \label{eq:copdef}
\end{equation}
Superscripts are therefore ordered from left to right in the symbol.
For a function of fixed weight, the ordinary maximal-length symbol
$\mathcal S$ does not detect positive-weight constants, including
zeta values, or their products with lower-weight functions.
The coproduct of the full function retains these terms~\cite{DGR,Duhr}.
For a weight-six symbol $S$, we also use
\begin{equation}
 V_k(S)=\operatorname{span}_{\mathbb Q}
 \{\text{weight-$k$ left factors of }\Delta_{k,1,\ldots,1}S\}.
 \label{eq:Vk}
\end{equation}
We use $V_k$ for left-factor spaces and state right-final-entry
relations separately.

\section{Bootstrapping the symbol}
\label{sec:symbol}
We reconstruct the symbol from a general six-letter ansatz, recursive
branch-cut conditions and the independently computed collinear limit.
The weak adjacency and single-final-entry properties observed in
Ref.~\cite{GMR} are supplemented by four families of triple-final-entry
relations extracted from that result. Together they select a unique
symbol. Table~\ref{tab:symbol} shows the number of parameters remaining after
each stage.

\subsection{The ansatz and branch-cut conditions}
Our ansatz is a pure weight-six multiple polylogarithm (MPL) with
rational coefficients and alphabet~\eqref{eq:alphabet}. Its symbol must be
integrable and invariant under $S_3$. At fixed total virtuality, the
physical thresholds are at $s_{ij}=0$, so the allowed first entries
are $u,v,w$. Period-dependent monodromies are treated in the function reconstruction. We exclude
adjacent distinct complementary letters, such as
$\bar u\otimes\bar v$, in the first four pairs of slots, but not in
the final pair. Writing $S=\mathcal S(F)$, we also impose the weak
single-final-entry relation~\cite{GMR}:
\begin{equation}
 \sum_{a\in\mathcal A} S^a=0.
 \label{eq:weaklast}
\end{equation}

We implement the branch-cut constraints recursively as in
Ref.~\cite{DMW}. For each left factor $g$ in the coproduct, let
$g^{\bar u}$ be the coefficient of $d\log(1-u)$ in $dg$, and require
\begin{equation}
 \operatorname{LP}_{\delta\to0}
 g^{\bar u}(1-\delta,\delta z,\delta(1-z))=0.
 \label{eq:pseudothresh}
\end{equation}
The leading-power operation retains the full coefficient of
$\delta^0$, including every power of $\log\delta$ and its dependence on
$z$. We impose this condition for every $0<z<1$, at symbol level,
on left factors of weights two through five. Together with weak
adjacency and~\eqref{eq:weaklast}, these recursive conditions define the
candidate space used below.

The three-loop splitting amplitude~\cite{Splitting}, with conventions
and analytic continuation matched to~\eqref{eq:norm}, supplies the
collinear boundary. The recursive branch conditions leave 101 free
coefficients, and collinear matching leaves 45. The final-entry equations
then select a unique symbol, as shown in Table~\ref{tab:symbol}.

\begin{table}[ht]
\centering
\begin{tabular}{lr}
\toprule
Conditions imposed & Remaining parameters\\
\midrule
First entries, $S_3$, weakened adjacency, integrability &261\\
Weak single-final entry~\eqref{eq:weaklast} &180\\
Lower-coproduct branch regularity &101\\
Collinear symbol &45\\
$T_1$ and its $S_3$ images &27\\
$T_2$ and its $S_3$ images &18\\
$T_3$ and its $S_3$ images &3\\
$T_4$ and its $S_3$ images &0\\
\bottomrule
\end{tabular}
\caption{Number of free rational coefficients after each set of
conditions. Integrability is imposed exactly as a rational-function
identity. The final row corresponds to a unique solution.}
\label{tab:symbol}
\end{table}

\subsection{Relations among the last three entries}
The four additional families involve at most four terms each:
\begin{align}
 T_1[F]&=F^{u,u,\bar v}+F^{w,u,\bar v}-F^{\bar v,\bar v,u},\nonumber\\
 T_2[F]&=F^{u,v,u}-F^{u,\bar u,\bar w}-F^{v,u,v}
                       +F^{v,\bar v,\bar w},\nonumber\\
 T_3[F]&=F^{u,u,u}+2F^{u,u,\bar u}-F^{u,\bar u,u}+F^{\bar u,u,u},\nonumber\\
 T_4[F]&=F^{u,u,\bar u}+2F^{u,\bar u,u}+F^{u,\bar u,\bar u}
                       -F^{\bar u,u,u}.
 \label{eq:T}
\end{align}
We impose $T_i[S]=0$ together with all $S_3$ images of these equations. These relations differ
from the planar final-entry equations~\cite{DMW,EightLoops}. The same
operators act on full-function coproducts in Section~\ref{sec:uplift}.

With the other conditions held fixed, omitting $T_1$, $T_2$, $T_3$, or
$T_4$ leaves 12, 4, 3, or 3 free parameters, respectively.
Appendix~\ref{sec:fourfinal} gives an alternative reconstruction using
four-final-entry equations.

The resulting symbol agrees exactly with Ref.~\cite{GMR}. Starting
from the 40-term collinear symbol, the linear system gives 4506 nonzero
words with coefficients $\{\pm1,\pm2,\pm3,\pm4\}$ in the
normalization~\eqref{eq:norm}. The soft limit, triple discontinuities,
and diagonal and high-energy limits provide additional checks.
Neither the small coefficients nor these additional limits are imposed
in the symbol reconstruction.

The SLC result does not belong to the planar symbol space
$\mathcal M_6$ of Ref.~\cite{DMW}: it contains 468 words with adjacent
distinct complementary letters, all in the last two positions. Its weight-five left factors do belong to the larger form-factor space
$\mathcal M_5$ of Ref.~\cite{DMW}, which should be distinguished from
the smaller space actually used by planar coproducts:
\begin{equation}
 \Delta_{5,1}\mathcal S(F)\in\mathcal M_5^{\rm symb}
          \otimes\operatorname{span}_{\mathbb Q}\mathcal A,
 \qquad \mathcal S(F)\notin\mathcal M_6^{\rm symb}.
 \label{eq:extension}
\end{equation}
The appropriate space for $F$ is therefore an integrable one-entry
extension of $\mathcal M_5$.

\section{The complete three-loop function in an \texorpdfstring{$A_2$}{A2} alphabet}
\label{sec:uplift}
We now give an expression for the full finite remainder, including
all terms invisible to the ordinary symbol. Summing a single
five-letter function over the six permutations gives
\begin{equation}
 \boxed{F(u,v,w)=\operatorname{Sym}\Phi_6^{A_2}(u,v),\qquad
 \Phi_6^{A_2}(u,v)=A_6(u)+\frac12B_6(u,v)+P_6(u,v).}
 \label{eq:master-explicit}
\end{equation}
The 35 weight-six GPLs in $A_6+B_6/2$ form the seed's chosen non-product
representative. The function $P_6$ contains the lower-weight products
and period terms. Every term in this expression belongs to the same
five-letter algebra.

\subsection{A five-letter \texorpdfstring{$A_2$}{A2} expression}
\label{sec:A2}
Throughout this section $w=1-u-v$, with $u,v,w>0$. We use the
six-term sum
\begin{equation}
 \operatorname{Sym}X=X(u,v)+X(v,u)+X(u,w)+X(w,u)+X(v,w)+X(w,v),
 \label{eq:sym-sum}
\end{equation}
and the GPL convention~\cite{DGR}
\begin{equation}
 G(a_1,\ldots,a_n;t)=\int_0^t\frac{ds}{s-a_1}G(a_2,\ldots,a_n;s),
 \qquad G(0^n;t)=\frac{\log^n t}{n!},\quad G(\varnothing;t)=1.
 \label{eq:GPL-convention}
\end{equation}
Define
\begin{equation}
\begin{gathered}
 U_{a_1\cdots a_n}=G(a_1,\ldots,a_n;u),\qquad a_i\in\{0,1\},\\
 V_{a_1\cdots a_n}=G(\rho_{a_1},\ldots,\rho_{a_n};v),\qquad
 (\rho_0,\rho_1,\rho_2)=(0,1-u,-u),\qquad a_i\in\{0,1,2\}.
\end{gathered}
\label{eq:UV-definitions}
\end{equation}
We use the same notation in Appendix~\ref{app:merged}. Setting
$x=v/(1-u)$ and $z=-u/(1-u)$ rescales the $V$ letters to
$\{0,1,z\}$. Since the GPL words below end in a nonzero letter, no
additional logarithms arise from the tangential basepoint.
The function $A_6(u)$ has seven GPL terms, while $B_6(u,v)$ has 28:
\begin{subequations}\label{eq:weight-six-core}
\begin{align}
% BEGIN_CORE A
A_6={}&-7U_{000001} -13U_{000011} -4U_{001011} -U_{001101} +69U_{001111} +3U_{010111} +53U_{011111}
% END_CORE A
\label{eq:A6}\\[4pt]
% BEGIN_CORE B
B_6={}&4V_{000121} +16V_{001121} -3V_{001212} -12V_{001221} +12V_{002011} +6V_{002101} -11V_{002121} +18V_{002212}\nonumber\\
&{}+12V_{002221} -2V_{010212} +2V_{011021} +28V_{011121} +16V_{011211} -14V_{011221} -2V_{012021} +V_{012112}\nonumber\\
&{}-10V_{012121} -V_{012122} -26V_{012211} -10V_{020211} +6V_{020212} +4V_{020221} -12V_{021111} +2V_{021112}\nonumber\\
&{}-25V_{021121} +23V_{022121} +64V_{022211} +15V_{111222}
% END_CORE B
\label{eq:B6}
\end{align}
\end{subequations}

Equations~\eqref{eq:A6} and~\eqref{eq:B6}, together with $P_6$
in Appendix~\ref{app:merged}, give the complete finite remainder
in~\eqref{eq:master-explicit}.

All GPLs appearing in the formula have letters $\{0,1\}$ when
the endpoint is $u$, and $\{0,1-u,-u\}$ when the endpoint is $v$.
Consequently, their differentials involve only
\begin{equation}
 \boxed{\mathcal A_{\rm seed}=\{u,v,w,1-u,1-w\}.}
 \label{eq:seed-alphabet}
\end{equation}
In the coordinates above,
\[
 \{x,1-x,z,1-z,x-z\}
 =\left\{\frac v{1-u},\frac w{1-u},-\frac u{1-u},
 \frac1{1-u},\frac{1-w}{1-u}\right\}.
\]
These are the five letters of the $A_2$ (or $M_{0,5}$)
iterated-integral alphabet~\cite{Brown,PSSV}. Equivalently, with
$r=u/v$ and $s=w/v$, one may use
$\{r,s,1+r,1+s,1+r+s\}$. The sixth letter of the full form-factor
alphabet, $1-v$, enters when we permute the arguments.
The $A_2$ description applies to the seed's iterated-integral alphabet;
the final-entry equations constrain the full permutation sum.

The period terms also belong to this algebra. For example,
$U_{01}+\zeta_2$ occurs as a coefficient of lower-weight functions in
$P_6$. Keeping such combinations together incorporates the zeta-value
terms directly into the product part of the seed.

\subsection{Determining the terms beyond the symbol}
We integrate the symbol along a regularized Chen path from the
tangential boundary $v=0$, with $v$ as the integration variable.
This produces an expression $F_0$ with 1495 terms in fibrated
GPLs~\cite{Brown,DGR}. In an ansatz of pure MPLs with multiple zeta values (MZVs) as
constants, the terms not determined by the symbol can be written as
\begin{align}
 F=F_0&+\zeta_2 f_4+\zeta_3 f_3+\zeta_2^2 f_2
      +\zeta_5 f_1+\zeta_2\zeta_3g_1\nonumber\\
      &+c_{222}\zeta_2^3+c_{33}\zeta_3^2.
 \label{eq:uplift}
\end{align}
The fibrated GPL ansatz has 658 rational coefficients; here the
subscripts on $f_k$ denote transcendental weight. Under permutations and analytic continuation around nonphysical branch
loci, $F_0$ acquires period terms that must cancel against the completion. Symmetry and monodromy conditions therefore
act on the full sum~\eqref{eq:uplift}, rather than on each correction
separately. For monodromy around $v=1$, we set $u=-1/t+i0$ and impose
the vanishing of the $t^0$ coefficient. The coefficients of the next eight powers of $t$ also vanish in the
reconstructed function.

For the collinear boundary we use the independently computed
three-loop splitting amplitude~\cite{Splitting}, matching its color
normalization and analytic continuation to~\eqref{eq:norm}. The soft
endpoint of this boundary, together with the soft current for two hard
partons~\cite{Soft3}, gives
\begin{equation}
 F_{\rm soft}=C_5\log\frac{uv}{w}
              -\frac{235}{8}\zeta_6-6\zeta_3^2+o(1)
 \label{eq:soft}
\end{equation}
as $u=\delta z$, $v=\delta(1-z)$ and $w=1-\delta$, with
$\delta\to0$ and $0<z<1$.
The function-level single-final-entry equation is
\begin{equation}
 \sum_{a\in\mathcal A}F^a=C_5.
 \label{eq:lastsource}
\end{equation}
The same constant $C_5$ governs the scale derivative in
\eqref{eq:scale}. Equation~\eqref{eq:lastsource}, however, acts on
kinematic-ratio coproducts at fixed $Q^2$ and is an empirical
final-entry input, separate from infrared evolution.

\begin{table}[ht]
\centering
\begin{tabular}{lr}
\toprule
Successive conditions & Free coefficients\\
\midrule
Allow all terms beyond the symbol~\eqref{eq:uplift} &658\\
Six-letter alphabet and $S_3$ symmetry &119\\
Independent collinear boundary &85\\
Soft boundary~\eqref{eq:soft} &69\\
Nonphysical-monodromy condition &9\\
Single-final entry~\eqref{eq:lastsource} &1\\
One full-function relation $T_1[F]=0$ &0\\
\bottomrule
\end{tabular}
\caption{Free rational coefficients in the function completion.
The leading monodromy coefficient is imposed in the reconstruction;
eight subsequent coefficients provide checks on the solution.}
\label{tab:uplift}
\end{table}

\subsection{The remaining freedom and the two-loop planar remainder}
\label{sec:ambiguity}
The first six rows of Table~\ref{tab:uplift} leave a one-dimensional
homogeneous solution space. Solving for its generator, we identify it
with $\zeta_2/8$ times the independently known two-loop planar
remainder~\cite{Form2,DMW}:
\begin{equation}
 \boxed{F_{\rm pre}=F_{\rm part}+cK_6,\qquad
 K_6=\frac{\zeta_2}{8}R_{\mathrm{LC}}^{(2)}.}
 \label{eq:planar-ambiguity}
\end{equation}
This identification holds at function level. The generator $K_6$
vanishes in the collinear and soft limits and obeys the homogeneous
single-final-entry equation, accounting for the remaining freedom.

The triple-final-entry operators act on this direction as follows:
\begin{equation}
 \boxed{\begin{aligned}
 T_1[K_6]&=\tfrac12\zeta_2\log v,& T_2[K_6]&=0,\\
 T_3[K_6]&=-\tfrac12\zeta_2\log u,&T_4[K_6]&=\tfrac12\zeta_2\log u.
 \end{aligned}}
 \label{eq:kernelimages}
\end{equation}
In particular, $T_1[F]=0$ determines the last coefficient.
With our choice of particular solution,
$T_1[F_{\rm part}]=\zeta_2\log v/10$, so
\begin{equation}
 F=F_{\rm part}-\frac{\zeta_2}{40}R_{\mathrm{LC}}^{(2)}.
 \label{eq:particularanswer}
\end{equation}
The coefficient in~\eqref{eq:particularanswer} depends on the choice of
$F_{\rm part}$, while the homogeneous direction is fixed. Either $T_3$ or
$T_4$ can replace $T_1$; $T_2$ annihilates the ambiguity. A logarithmic
source relation provides another determination in
Section~\ref{sec:lasttwo}.

The reconstructed function agrees exactly with Ref.~\cite{GMR},
passes the remaining monodromy checks and satisfies all permuted
triple-final-entry relations. Its 1763-term fibrated GPL expression is reorganized into
the complete $A_2$ seed in~\eqref{eq:master-explicit}.

\section{Planar comparison and final-entry structure}
\label{sec:comparison}\label{sec:finalstructure}
The compact expression of Section~\ref{sec:uplift} combines small
symbol coefficients with function content that is distinct from the
planar result. Appendix~\ref{app:coproduct} gives the coproduct spaces and their
explicit bases; Appendix~\ref{app:finalentries} gives the final-entry
relations and their use in the bootstrap.

\subsection{Shared functions and new nonclassical content}
Table~\ref{tab:comparison} compares the three symbols in the same
alphabet~\eqref{eq:alphabet}. In each column we clear denominators and
divide by the greatest common divisor of the integer coefficients.
This removes the overall numerical normalization, while retaining the
different finite-remainder conventions of $\mathcal E^{(3)}$ and
$R^{(3)}$.

\begin{table}[ht]
\centering
\begin{tabular}{lrrr}
\toprule
 &$\mathcal E^{(3)}$&$R^{(3)}$&$F$\\
\midrule
Nonzero symbol words &4152&4770&4506\\
Number of distinct absolute coefficients &18&15&4\\
Largest absolute coefficient &60&48&4\\
\bottomrule
\end{tabular}
\caption{Symbol support and integer coefficients in a fixed alphabet,
with each symbol normalized as described in the text. The SLC absolute
coefficients are $1,2,3,4$.}
\label{tab:comparison}
\end{table}

The SLC symbol has a comparable number of terms but substantially
smaller coefficients. This arithmetic simplicity does not make $F$ a multiple
of the planar remainder: even modulo lower-weight products and
zero-symbol terms, the two functions are independent. The two planar
normalizations, by contrast, differ only by products and periods.
Appendix~\ref{app:quotient} gives the comparison and a short certificate.

The low-weight coproducts show where the new content appears. Let
$\pi_k$ project weight-$k$ symbols modulo shuffle products of
positive-weight symbols.
At weight three, all three projected left-factor spaces coincide:
\begin{equation}
 \pi_3 V_3(\mathcal S(\mathcal E^{(3)}))
 =\pi_3 V_3(\mathcal S(R^{(3)}))
 =\pi_3 V_3(\mathcal S(F)),\qquad \dim=8.
 \label{eq:V3modproducts}
\end{equation}
Before this projection, the three extra SLC directions relative to
$R^{(3)}$ can be chosen as
$\mathcal S(\log u_i\,\operatorname{Li}_2(1-u_i))$, for $u_i=u,v,w$.
At weight four, the SLC result has additional nonclassical content. The
antisymmetric coproduct $\delta_{2,2}$, with products removed from both
weight-two factors, gives
\begin{equation}
 \begin{gathered}
 \delta_{2,2}V_4(\mathcal S(\mathcal E^{(3)}))
 =\delta_{2,2}V_4(\mathcal S(R^{(3)}))
 \subset\delta_{2,2}V_4(\mathcal S(F)),\\
 \dim\delta_{2,2}V_4(\mathcal S(R^{(3)}))=5,\qquad
 \dim\delta_{2,2}V_4(\mathcal S(F))=9.
 \end{gathered}
 \label{eq:V4cobrackets}
\end{equation}
The four additional directions have a basis of dilogarithmic wedges
and arise in coproducts of the deviations from planar single-final-entry
conditions. Together with the five planar directions, they span the
full nonclassical image of the weight-four space
$\mathcal M_4^{\rm symb}$. Explicit bases and their relation to final
entries are given in Appendix~\ref{app:weightfour}.

The connection to the planar sector also extends beyond symbols.
The
weight-three coproducts of $R_{\rm LC}^{(2)}$ lie in the corresponding
space for $F$ after a fixed $\zeta_3$ shift, as described in
Appendix~\ref{sec:planarperiods}. This complements the appearance of
$\zeta_2R_{\rm LC}^{(2)}$ as the last ambiguity in the full-function
reconstruction, Eq.~\eqref{eq:planar-ambiguity}.

\subsection{Last-two adjacency and the function completion}
\label{sec:lasttwo}
Strict complementary-letter adjacency holds for the planar BDS-like
quantity $\mathcal E^{(3)}$, but not for $R^{(3)}$. For the SLC result,
all violations are confined to the last pair. The three independent
functions $A_{ij}=F^{\bar i,\bar j}=F^{\bar j,\bar i}$, with $i\ne j$,
encode these violations. Although the SLC coproduct has the additional
nonclassical content in~\eqref{eq:V4cobrackets}, each $A_{ij}$ is a
classical weight-four function in a five-letter $A_2$ alphabet. The
complete expression for $A_{uv}$, including period terms, is given in
Appendix~\ref{app:classical}; the other two follow by permutation.

These functions have simple logarithmic and one-mass-box coproducts.
For example, $A_{uv}^{u}+A_{uv}^{w}
=\tfrac12\log^2v\log(u/w)$, and
\begin{equation}
 \boxed{F^{w,w,\bar u,\bar v}=b_\Box(u,v).}
 \label{eq:box-descent}
\end{equation}
Here $b_\Box$ is the finite one-mass-box function~\cite{BDDK}, written
explicitly in~\eqref{eq:defectsources}. The logarithmic source can
replace one family of triple-final constraints in the symbol
bootstrap and also fixes the remaining
$\zeta_2R_{\rm LC}^{(2)}/8$ ambiguity at function level. The derivation
of the associated differential equation and the alternative
reconstruction are given in Appendix~\ref{app:sources}.

The adjacency violation cannot be removed by a finite subtraction
that is a constant-coefficient linear combination of products of
lower-weight pure functions, each with physical first entries. The
corresponding weight-four defect has a nonzero component modulo products;
Appendix~\ref{sec:subtraction} proves the obstruction. This differs from
the two-loop planar case, where a product of one-loop functions
restores adjacency.

There are also four families of two-term four-final relations. Together with $T_1$, they
provide an alternative symbol reconstruction, but all annihilate the
last function-level ambiguity $K_6$. The triple- and four-final
equations therefore play different roles in reconstructing the symbol
and fixing the period terms. Their explicit forms and the resulting parameter
counts are collected in Appendix~\ref{sec:fourfinal}.

\section{Towards a four-loop bootstrap}
\label{sec:fourloops}
The three-loop constraints also define a possible starting point at
four loops. We apply them to the connected finite correction and obtain
a weight-eight symbol space of exact rational dimension 200. Its
collinear maps specify boundary data that would test the ansatz and
select a candidate symbol. The known full-color integrand~\cite{LYZ}
can be used to compute these data.

\subsection{The four-loop finite term}
In the pole-only scheme of Section~\ref{sec:setup}, write
\begin{equation}
 H=P(a)+\nu[a^3f_3+a^4f_4+\cdots],\qquad
 P=1+ap_1+\cdots,\qquad \nu=N_c^{-2}.
 \label{eq:Hexpand}
\end{equation}
Here $a=a_s$, $p_1$ is the one-loop planar contribution, and
$f_3=48F$ at the scale in~\eqref{eq:norm}.
At four loops, we consider the coefficient of $a^4/N_c^2$ in $\log H$:
\begin{equation}
 K_4=[\log H]_{a^4\nu}=f_4-p_1f_3.
 \label{eq:K4}
\end{equation}
In the planar BDS-like convention, the known product $p_1f_3$
violates adjacency in internal positions and gives nonzero values
for the triple-final-entry combinations in~\eqref{eq:T}. Thus imposing homogeneous versions of the
three-loop equations on $K_4$ is different from imposing them on
$f_4$. For example,
\begin{equation}
 T_i[\mathcal S(f_4)]=T_i[\mathcal S(p_1f_3)]
 \quad\text{if}\quad T_i[\mathcal S(K_4)]=0.
 \label{eq:fourloop-source}
\end{equation}
The connected coefficient $K_4$ is invariant under a common
color-independent finite normalization. We keep color-dependent finite
terms fixed by the pole-only prescription and test homogeneous
final-entry relations for $K_4$.

We now define the weight-eight symbol spaces under these
assumptions. Let $\mathcal W_8$ be the $\mathbb Q$-vector space of
integrable, $S_3$-invariant weight-eight symbols with
alphabet~\eqref{eq:alphabet} and first entries in $\{u,v,w\}$.
We impose the weak single-final-entry relation~\eqref{eq:weaklast}
and exclude $\bar i\otimes\bar j$ for $i\ne j$ in slots $(r,r+1)$
with $r=1,\ldots,6$, allowing only the final pair to violate adjacency.
The recursive branch conditions~\eqref{eq:pseudothresh} are imposed
on all left factors of weights two through seven. Period completion
is a subsequent step.
Imposing the $T_i$ relations gives
\begin{equation}
 \mathcal V_{8,T}
 =\{S\in\mathcal W_8:T_i[S]=0,\ i=1,\ldots,4\}.
 \label{eq:V8T-def}
\end{equation}
We include all permutations of the four $T_i$ relations in
\eqref{eq:T}, but do not impose the four-final-entry relations
of Appendix~\ref{sec:fourfinal}.

In the timelike collinear limit, set $u=\delta$, $v=z$ and
$w=1-z-\delta$, at fixed $q^2$ and subtraction scale. The leading
collinear limit of a symbol, including every power of
$L=\log\delta$, defines the map $\mathfrak C$. We choose a
subtraction scheme compatible with collinear factorization:
$H_3\to H_{\rm Sp}H_2$, where $H_{\rm Sp}$ is obtained from the
splitting amplitude by using $Z_3/Z_2$ in the corresponding color
channel. For $\beta=0$, infrared evolution and the assumed
transcendental weights allow the form~\cite{IR}
\begin{equation}
 K_4\big|_{\rm coll}=b_4(z)+c_6L^2+
 [d_6\log z+e_6\log(1-z)+c_7]L.
 \label{eq:fourloop-collinear-scale}
\end{equation}
The subscripts on the constants indicate their transcendental weights.
The function $b_4(z)$ contains the unknown finite splitting contribution
and a two-point constant. The coefficients of the explicit logarithms
have positive transcendental weight, so their ordinary weight-eight
symbols vanish. This gives the additional symbol-space restriction
\begin{equation}
 \mathcal V_{8,\rm IR}
 =\{S\in\mathcal V_{8,T}:
 \mathfrak C S\in\operatorname{span}_{\mathbb Q}\{z,1-z\}^{\otimes8}\}.
 \label{eq:V8IR-def}
\end{equation}
Our four-loop assumption can then be stated precisely as
\begin{equation}
 \boxed{\mathcal S(K_4)\ \stackrel{\rm conj.}{\in}\
              \mathcal V_{8,\rm IR}.}
 \label{eq:fourloop-space-conjecture}
\end{equation}
Purity, the six-letter alphabet, weak adjacency, recursive branch
conditions and the homogeneous $T_i$ equations are the structural
assumptions in~\eqref{eq:fourloop-space-conjecture}. Within this ansatz,
infrared evolution supplies the collinear scale restriction. The
remaining finite boundary $b_4(z)$ is dynamical input.

\subsection{Dimensions of the weight-eight symbol spaces}
Table~\ref{tab:fourloop-candidate} gives the dimensions obtained by
successively imposing these conditions. We compute exact rational
kernel bases and use modular rank calculations to establish their
completeness.

\begin{table}[ht]
\centering\small
\begin{tabular}{p{0.70\linewidth}r}
\toprule
Cumulative conditions & Dimension over $\mathbb Q$\\
\midrule
Initial space $\mathcal W_8$ &844\\
Add $T_1$ and its permutations &665\\
Add $T_2$ and its permutations &554\\
Add $T_3$ and its permutations &422\\
Add $T_4$: $\mathcal V_{8,T}$ &290\\
Collinear restriction: $\mathcal V_{8,\rm IR}$ &200\\
Kernel of the leading collinear limit &81\\
\bottomrule
\end{tabular}
\caption{Exact dimensions of the candidate weight-eight spaces.
The final row is the homogeneous kernel of the collinear map.}
\label{tab:fourloop-candidate}
\end{table}

For each constraint matrix $A$, an integer matrix $B$ gives a full set
of independent kernel vectors with $AB=0$. A matching modular rank
bound proves completeness over $\mathbb Q$.

\subsection{Further collinear information}
\label{sec:probes}
The collinear map $\mathfrak C$ has rank 119 on
$\mathcal V_{8,\rm IR}$ and an 81-dimensional kernel. An independently computed boundary tests whether the physical data
lie in this image. If they do, matching the boundary leaves an affine
family with 81 parameters.
Until that boundary is known, all 200 coefficients remain free.
The leading soft-symbol limit gives no further restriction on the kernel.

Collinear factorization expresses the unknown boundary in terms of
finite splitting amplitudes and the two-point form factor. Write
$H_{\rm Sp}=S_{\rm LC}+\nu(a^3\sigma_3+a^4\sigma_4+\cdots)$, with
$S_{\rm LC}=1+as_1+\cdots$, and denote the four-loop SLC coefficient
of the finite two-point form factor by $d_4$. Expanding
$H_3\to H_{\rm Sp}H_2$ gives
\begin{equation}
 K_4\big|_{\rm coll}=\sigma_4-s_1\sigma_3+d_4.
 \label{eq:connected-splitting}
\end{equation}
At fixed total virtuality and subtraction scale, $d_4$ is independent
of $z$ and has no ordinary weight-eight symbol. The required $z$ dependence is supplied by the connected four-loop
finite splitting amplitude. Lower-loop splitting and four-loop Sudakov
results~\cite{Splitting,Sudakov4} supply the remaining factors.
Expanding the known integrand in collinear regions before integration
provides a route to this single-variable input~\cite{Splitting,LYZ}.

To distinguish the remaining 81 directions, consider the boundary
coproducts of a symbol $S\in\ker(\mathfrak C|_{\mathcal V_{8,\rm IR}})$:
\begin{equation}
 \mathcal B_{\rm sym}(S)=\bigl(\mathfrak C S^w,
                    \mathfrak C S^{\bar u},
                    \mathfrak C S^{\bar w}\bigr).
 \label{eq:Bprobe}
\end{equation}
The three entries of this map are the leading collinear symbols
of weight-seven coproducts. Their relevance can be seen from the
differential equation for a function $X$ with symbol $S$:
\begin{equation}
 \partial_\delta X=\frac{X^u}{\delta}
 -\frac{X^w}{1-z-\delta}
 -\frac{X^{\bar u}}{1-\delta}
 +\frac{X^{\bar w}}{z+\delta}.
 \label{eq:radial}
\end{equation}
Extracting the three limits in $\mathcal B_{\rm sym}$ from a
near-collinear expansion requires separating the last three terms
from $X^u/\delta$. In particular, the $O(\delta)$ term of $X^u$
also contributes at order $\delta^0$ to the radial derivative.

As a map on the 81-dimensional kernel,
$\mathcal B_{\rm sym}$ is injective. The ranks below include each
indicated power of $L$ and all higher powers:
\begin{equation}
\begin{array}{c|rrrrrr}
 \text{Included powers of }L\text{, starting at}&5&4&3&2&1&0\\\hline
 \text{Rank}&1&3&13&36&78&81
\end{array}.
\label{eq:proberank}
\end{equation}
\Needspace{5\baselineskip}
In these coproduct limits, the $L^6$ and $L^7$ coefficients vanish.
Positive powers of $L$ distinguish 78 directions, and the constant
terms distinguish the remaining three. Thus a compatible leading
collinear symbol and the three coproduct limits in~\eqref{eq:Bprobe}
select at most one element of $\mathcal V_{8,\rm IR}$. Computing these
physical boundary values is the next step in the four-loop bootstrap.

\section{Conclusions}
\label{sec:conclusions}
The complete three-loop SLC finite remainder can be written as a
permutation sum of a single $A_2$-type function. The non-product part
of the seed contains 35 weight-six GPLs, while all products and period
terms use the same five-letter alphabet. The established maximal-transcendentality
correspondence extends this simplicity to the weight-six SLC sector of
Higgs-plus-three-gluon amplitudes in QCD.

The bootstrap determines both the symbol and the zeta-value terms
needed to complete the function. The remaining homogeneous ambiguity is
$\zeta_2R_{\rm LC}^{(2)}/8$. It is fixed by either a triple-final-entry
relation or the logarithmic source for the last-two defect, whereas
the four-final-entry relations annihilate this direction. The coproduct
comparison makes the relation to the planar sector more precise.
The weight-three left factors coincide modulo products. At weight four,
the SLC image contains four directions absent from the planar result.
Together with the five planar directions, they span the full
nine-dimensional nonclassical image of $\mathcal M_4^{\rm symb}$.

A field-theoretic derivation of the final-entry relations should
account for both the classical box and logarithmic sources and the
four additional nonclassical directions described in Appendix~\ref{app:weightfour}.
Their appearance in the single-final-entry deviations identifies
specific coproduct combinations to study using cuts or differential
equations for the physical combination of integrals.

At four loops, the proposed conditions define a 200-dimensional symbol
space. A connected-splitting boundary compatible with its collinear
image leaves an 81-dimensional affine family. The known
integrand~\cite{LYZ} offers a route to this boundary and to further
coproduct data. The companion two-loop four-point calculation~\cite{FF4SLC} provides
a finite-color result at higher multiplicity, building on planar
bootstraps and one-mass five-point integrals~\cite{FF4MHV,FF4Function,FF4Analytic,FF4NMHV,OneMass2}.
Comparing the two multiplicities should help identify which
function-space restrictions survive beyond planar color ordering.

An extension of antipodal duality~\cite{Antipodal} would require a broader
target than the standard hexagon first-entry space: reversing the SLC
symbol gives five independent first entries, compared with three for
the hexagon amplitude. At four loops, the maximal-transcendentality
relation
\begin{equation}
 \mathsf W_8 h^{(4),Hggg}_{+++,\mathrm{SLC}}
 \stackrel{?}{=}
 \mathsf W_8 h^{(4),Hggg}_{-++,\mathrm{SLC}}
 \stackrel{?}{=}f_4
 \label{eq:Higgs4-conjecture}
\end{equation}
is distinct from the space conjecture
in~\eqref{eq:fourloop-space-conjecture}.
If this correspondence persists at four loops, the finite-color
bootstrap would provide the common highest-weight SLC contribution
to Higgs-plus-three-gluon amplitudes.

\section*{Acknowledgements}
The author thanks Westlake University for hospitality.
OpenAI's GPT-6 Astra model was used under the
author's direction to assist with computational checks and manuscript
preparation. The author reviewed and verified all AI-assisted material
and scientific conclusions and takes full responsibility for the content.
The work is supported by the National Natural Science Foundation of China
under Grant Nos.~12225510 and 12247103, and by the New Cornerstone Science
Foundation.

\clearpage
\appendix
\begingroup\small
\allowdisplaybreaks[4]
\setlength{\jot}{0pt}
\raggedbottom
\setlength{\abovedisplayskip}{3pt}
\setlength{\belowdisplayskip}{3pt}
\setlength{\abovedisplayshortskip}{3pt}
\setlength{\belowdisplayshortskip}{3pt}
\setlength{\parskip}{1pt}
\section{Products and zeta-value terms in the complete expression}
\label{app:merged}
We use the $U,V$ notation of~\eqref{eq:UV-definitions} at $Q^2=1$.
The indices $0,1$ label the $U$ letters $\{0,1\}$, and $0,1,2$
label the $V$ letters $\{0,1-u,-u\}$. In this appendix, set
\begin{equation}
 a=\log u,\quad b=\log(1-u),\quad c=\log v,\quad
 e=\log\frac{w}{1-u},\quad f=\log\frac{u+v}{u}.
 \label{eq:appendix-logs}
\end{equation}
Each term in $P_6$ has total transcendental weight six.
The functions $B_a,B_b,B_c,B_f$ have weight five. If
$H\in\mathcal H_k$, its coefficient $Q_H$ has weight $6-k$,
and $\ell_j$ has weight $j$. The complete expression is
\begin{equation}
 P_6=aB_a+bB_b+cB_c+fB_f+
 \sum_{H\in\mathcal H_4\cup\mathcal H_3\cup\mathcal H_2}H Q_H
 +\sum_{k=0}^{5}a^k\ell_{6-k}.
 \label{eq:merged-seed}
\end{equation}
Each set $\mathcal H_k$ contains the weight-$k$ GPLs whose
coefficients $Q_H$ are given below, with no repeated terms.
Every factor multiplies its entire displayed coefficient. No
further permutation sum or additive constant is understood.

\noindent\textit{Weight five.}
\begin{align*}
% BEGIN_COMPACT B_a
2\,B_a={}&-82\,U_{00001}-50\,U_{00011}-24\,U_{00101}-128\,U_{00111}-82\,U_{01011}+90\,U_{01111}+12\,V_{00001}-22\,V_{00011}\\*
&{}+11\,V_{00021}-7\,V_{00101}+34\,V_{00111}-28\,V_{00112}-34\,V_{00121}+24\,V_{00122}+6\,V_{00211}+4\,V_{00212}+44\,V_{00221}\\*
&{}+7\,V_{01011}-12\,V_{01012}-26\,V_{01021}+12\,V_{01022}-10\,V_{01102}-2\,V_{01112}+29\,V_{01121}+6\,V_{01202}+3\,V_{01211}\\*
&{}-3\,V_{01212}+6\,V_{02021}-V_{02111}+16\,V_{02112}-22\,V_{02121}-10\,V_{02122}-7\,V_{02211}+4\,V_{02212}+49\,V_{02221}
% END_COMPACT B_a
\\[1pt]
% BEGIN_COMPACT B_b
2\,B_b={}&22\,U_{00011}+11\,U_{00101}-16\,U_{00111}-7\,U_{01011}+24\,U_{01111}+4\,V_{00012}-18\,V_{00021}+16\,V_{00112}-12\,V_{00122}\\*
&{}+6\,V_{00201}+18\,V_{00211}-14\,V_{00212}+12\,V_{00222}+6\,V_{01021}+2\,V_{01102}+28\,V_{01112}+4\,V_{01121}-14\,V_{01122}\\*
&{}-2\,V_{01202}-12\,V_{01211}-9\,V_{01212}-26\,V_{01221}-12\,V_{02021}+4\,V_{02022}-28\,V_{02112}+4\,V_{02121}+V_{02122}\\*
&{}+18\,V_{02211}+23\,V_{02212}+70\,V_{02221}
% END_COMPACT B_b
\\[1pt]
% BEGIN_COMPACT B_c
2\,B_c={}&-3\,U_{00001}+233\,U_{00011}+100\,U_{00101}-71\,U_{00111}+4\,U_{01011}-75\,U_{01111}-5\,V_{00102}+5\,V_{00112}\\*
&{}+7\,V_{00121}+4\,V_{00122}+5\,V_{00201}-2\,V_{00212}-6\,V_{01012}-V_{01021}+2\,V_{01022}-5\,V_{01102}+19\,V_{01112}\\*
&{}+7\,V_{01121}+V_{01202}-3\,V_{01211}
% END_COMPACT B_c
\\[1pt]
% BEGIN_COMPACT B_f
2\,B_f={}&-12\,U_{00001}+12\,U_{00011}+3\,U_{00101}-24\,U_{00111}-13\,U_{01011}
% END_COMPACT B_f
\end{align*}

\noindent\textit{Weight four.}
\begin{align*}
% BEGIN_COMPACT Q_U0001
6\,Q_{U_{0001}}={}&-240\,U_{01}+162\,V_{01}-114\,V_{02}+231\,V_{12}+57\,a^{2}+81\,a\,c-96\,a\,f-57\,b^{2}+207\,b\,c-231\,b\,f\\*
&{}-63\,c^{2}-33\,c\,f-267\,e\,f-166\,\zeta_2
% END_COMPACT Q_U0001
\\[1pt]
% BEGIN_COMPACT Q_U0011
6\,Q_{U_{0011}}={}&6\,U_{01}+102\,V_{01}+102\,V_{02}-30\,V_{12}-15\,a^{2}+21\,a\,b+66\,a\,c+18\,a\,f-51\,b^{2}+150\,b\,c+27\,b\,f\\*
&{}-75\,c^{2}+12\,c\,f-3\,f^{2}-190\,\zeta_2
% END_COMPACT Q_U0011
\\[1pt]
% BEGIN_COMPACT Q_U0111
6\,Q_{U_{0111}}={}&228\,U_{01}+396\,V_{01}-144\,V_{02}-27\,V_{12}-87\,a^{2}-234\,a\,b+207\,a\,c+48\,a\,f-216\,b^{2}+267\,b\,c\\*
&{}+18\,b\,f-88\,\zeta_2
% END_COMPACT Q_U0111
\\[1pt]
% BEGIN_COMPACT Q_V0002
2\,Q_{V_{0001}}={}&-32\,U_{01}-12\,a^{2}+19\,a\,b-12\,a\,c-32\,\zeta_2
% END_COMPACT Q_V0002
\qquad % BEGIN_COMPACT Q_V0003
2\,Q_{V_{0002}}={}11\,a\,b-18\,b^{2}
% END_COMPACT Q_V0003
\\[1pt]
% BEGIN_COMPACT Q_V0022
2\,Q_{V_{0011}}={}&-18\,U_{01}-12\,a^{2}+31\,a\,b+35\,a\,c
% END_COMPACT Q_V0022
\\[1pt]
% BEGIN_COMPACT Q_V0023
2\,Q_{V_{0012}}={}&-32\,U_{01}+15\,a^{2}-5\,a\,b-9\,a\,c+7\,b\,c+3\,b\,f-25\,\zeta_2
% END_COMPACT Q_V0023
\\[1pt]
% BEGIN_COMPACT Q_V0032
Q_{V_{0021}}={}&-19\,U_{01}-3\,V_{01}-2\,a^{2}+16\,a\,b-9\,a\,c-5\,a\,f+9\,b^{2}+2\,b\,c+3\,b\,f-\tfrac{115}{6}\,\zeta_2
% END_COMPACT Q_V0032
\\[1pt]
% BEGIN_COMPACT Q_V0033
Q_{V_{0022}}={}&5\,U_{01}+10\,a\,b+7\,b^{2}-b\,c
% END_COMPACT Q_V0033
\qquad % BEGIN_COMPACT Q_V0203
2\,Q_{V_{0102}}={}-6\,U_{01}+8\,a^{2}-15\,a\,b-4\,a\,c+6\,b^{2}+3\,b\,f-9\,\zeta_2
% END_COMPACT Q_V0203
\\[1pt]
% BEGIN_COMPACT Q_V0223
2\,Q_{V_{0112}}={}&-15\,U_{01}-16\,a^{2}+30\,a\,b-5\,a\,c-4\,b^{2}+5\,b\,c-3\,b\,f
% END_COMPACT Q_V0223
\\[1pt]
% BEGIN_COMPACT Q_V0232
4\,Q_{V_{0121}}={}&-14\,U_{01}-11\,a^{2}-2\,a\,b+2\,a\,c-24\,b^{2}-8\,b\,c-6\,b\,f-2\,c^{2}
% END_COMPACT Q_V0232
\\[1pt]
% BEGIN_COMPACT Q_V0233
2\,Q_{V_{0122}}={}&-U_{01}+10\,a^{2}+2\,a\,c-13\,b^{2}
% END_COMPACT Q_V0233
\qquad % BEGIN_COMPACT Q_V0322
2\,Q_{V_{0211}}={}-65\,U_{01}+a^{2}+34\,a\,b-64\,a\,c-16\,a\,f+6\,b^{2}+2\,b\,c
% END_COMPACT Q_V0322
\\[1pt]
% BEGIN_COMPACT Q_V0323
4\,Q_{V_{0212}}={}&10\,U_{01}-8\,a^{2}-54\,a\,b-12\,a\,c+8\,a\,f+7\,b^{2}+2\,b\,c
% END_COMPACT Q_V0323
\\[1pt]
% BEGIN_COMPACT Q_V0332
2\,Q_{V_{0221}}={}&27\,U_{01}+3\,a^{2}-13\,a\,b-24\,a\,c+2\,a\,f+18\,b^{2}+4\,b\,c
% END_COMPACT Q_V0332
\qquad % BEGIN_COMPACT Q_V0333
2\,Q_{V_{0222}}={}49\,a\,b+38\,b^{2}
% END_COMPACT Q_V0333
\end{align*}

\noindent\textit{Weight three.}
\begin{align*}
% BEGIN_COMPACT Q_U001
Q_{U_{001}}={}&13\,U_{001}+U_{01}\,(17\,a-\tfrac{5}{2}\,b-\tfrac{65}{2}\,c+f)+V_{01}\,(6\,a+\tfrac{7}{2}\,b+\tfrac{11}{2}\,c)+V_{02}\,(18\,a-27\,b-2\,c)\\*
&{}+V_{12}\,(-14\,a+30\,b+\tfrac{7}{2}\,c+\tfrac{51}{2}\,e-\tfrac{11}{2}\,f)+\zeta_2\,(\tfrac{134}{3}\,a+\tfrac{17}{2}\,b+\tfrac{7}{2}\,f)-24\,\zeta_3-\tfrac{7}{4}\,a^{3}-\tfrac{3}{2}\,b\,a^{2}+\tfrac{3}{4}\,c\,a^{2}\\*
&{}+\tfrac{51}{4}\,f\,a^{2}-6\,a\,b^{2}-\tfrac{9}{2}\,a\,b\,c+16\,a\,b\,f-\tfrac{7}{4}\,a\,c^{2}+\tfrac{11}{2}\,a\,c\,f+18\,a\,e\,f+a\,f^{2}+\tfrac{3}{4}\,b^{3}-\tfrac{25}{4}\,c\,b^{2}-\tfrac{59}{4}\,f\,b^{2}\\*
&{}+12\,b\,c^{2}-4\,b\,c\,f-\tfrac{57}{2}\,b\,e\,f+4\,b\,f^{2}-\tfrac{5}{2}\,c\,e\,f-\tfrac{55}{4}\,f\,e^{2}+\tfrac{5}{2}\,e\,f^{2}
% END_COMPACT Q_U001
\\[1pt]
% BEGIN_COMPACT Q_U011
Q_{U_{011}}={}&\tfrac{3}{2}\,U_{001}+13\,U_{01}\,a+\tfrac{5}{2}\,U_{01}\,b-\tfrac{11}{2}\,U_{01}\,c+4\,U_{01}\,f-\tfrac{23}{2}\,U_{011}-38\,V_{01}\,a-\tfrac{99}{2}\,V_{01}\,b-\tfrac{11}{2}\,V_{01}\,c+9\,V_{02}\,a\\*
&{}+26\,V_{02}\,b+4\,V_{12}\,a+\tfrac{3}{2}\,V_{12}\,b-2\,V_{12}\,e-2\,V_{12}\,f-\tfrac{3}{4}\,a^{3}+\tfrac{41}{4}\,b\,a^{2}+\tfrac{23}{4}\,c\,a^{2}-\tfrac{17}{4}\,f\,a^{2}+17\,a\,b^{2}\\*
&{}-\tfrac{33}{2}\,a\,b\,c-\tfrac{3}{2}\,a\,b\,f-\tfrac{19}{4}\,a\,c^{2}-4\,a\,c\,f+a\,e\,f+\tfrac{79}{6}\,a\,\zeta_2+\tfrac{209}{12}\,b^{3}-\tfrac{103}{4}\,c\,b^{2}-\tfrac{1}{2}\,f\,b^{2}+\tfrac{11}{2}\,b\,c^{2}+\tfrac{3}{2}\,b\,c\,f\\*
&{}+2\,b\,e\,f+\tfrac{3}{2}\,b\,f^{2}-\tfrac{7}{2}\,b\,\zeta_2+c\,e\,f+f\,e^{2}+\tfrac{3}{2}\,e\,f^{2}+5\,f\,\zeta_2-\tfrac{9}{2}\,\zeta_3
% END_COMPACT Q_U011
\\[1pt]
% BEGIN_COMPACT Q_V002
24\,Q_{V_{001}}={}&-348\,U_{001}+U_{01}\,(372\,a-420\,b+588\,c-36\,f)-948\,U_{011}+108\,V_{01}\,a-36\,V_{02}\,b\\*
&{}+3\,a\,(41\,a^{2}-60\,a\,b+24\,a\,c+40\,a\,f+80\,b^{2}-44\,b\,c-12\,b\,f)+\zeta_2\,(220\,a-264\,b)+1416\,\zeta_3
% END_COMPACT Q_V002
\\[1pt]
% BEGIN_COMPACT Q_V003
12\,Q_{V_{002}}={}&-270\,U_{001}+360\,U_{01}\,a-18\,U_{01}\,b+48\,U_{01}\,c-18\,U_{01}\,f+78\,U_{011}+18\,V_{01}\,a-36\,V_{01}\,b-18\,V_{02}\,b\\*
&{}-54\,b\,a^{2}+60\,e\,a^{2}+75\,a\,b^{2}-54\,a\,b\,c-60\,a\,b\,f+108\,a\,\zeta_2+54\,b^{3}+9\,c\,b^{2}+18\,f\,b^{2}-62\,b\,\zeta_2
% END_COMPACT Q_V003
\\[1pt]
% BEGIN_COMPACT Q_V022
8\,Q_{V_{011}}={}&236\,U_{001}-20\,U_{01}\,a-20\,U_{01}\,b-12\,U_{01}\,f-36\,U_{011}+40\,V_{02}\,a-12\,V_{02}\,b-5\,a^{3}-12\,b\,a^{2}\\*
&{}+30\,c\,a^{2}+64\,f\,a^{2}+32\,a\,b^{2}-12\,a\,b\,f
% END_COMPACT Q_V022
\\[1pt]
% BEGIN_COMPACT Q_V023
4\,Q_{V_{012}}={}&16\,U_{001}-32\,U_{01}\,a-28\,U_{01}\,b-4\,U_{01}\,c+8\,U_{011}+6\,V_{001}+6\,V_{002}+4\,V_{01}\,a+6\,V_{01}\,b+2\,V_{01}\,c\\*
&{}+6\,V_{011}-6\,V_{02}\,a+13\,a^{3}+23\,b\,a^{2}+8\,c\,a^{2}-5\,f\,a^{2}-a\,b^{2}+6\,a\,b\,c+6\,a\,b\,f-10\,a\,\zeta_2-13\,b^{3}\\*
&{}-6\,c\,b^{2}-6\,f\,b^{2}-2\,b\,c^{2}-14\,b\,\zeta_2-2\,c\,\zeta_2+2\,\zeta_3
% END_COMPACT Q_V023
\\[1pt]
% BEGIN_COMPACT Q_V032
4\,Q_{V_{021}}={}&-108\,U_{001}+U_{01}\,(-8\,a-112\,b-8\,c-4\,e+4\,f)+100\,U_{011}+4\,V_{01}\,a-4\,V_{02}\,b-2\,V_{021}\\*
&{}+\zeta_2\,(-122\,a+22\,b)-10\,a^{3}-2\,b\,a^{2}-4\,c\,a^{2}+3\,e\,a^{2}-8\,f\,a^{2}+66\,a\,b^{2}-120\,a\,b\,c+4\,a\,b\,e\\*
&{}-28\,a\,b\,f+8\,a\,c\,f+10\,b^{3}+4\,c\,b^{2}+2\,e\,b^{2}-2\,b\,c\,f
% END_COMPACT Q_V032
\\[1pt]
% BEGIN_COMPACT Q_V033
4\,Q_{V_{022}}={}&110\,U_{01}\,a+68\,U_{01}\,b+32\,U_{011}+18\,b\,a^{2}+32\,e\,a^{2}-19\,a\,b^{2}-52\,a\,b\,c+4\,a\,b\,f+134\,a\,\zeta_2\\*
&{}+18\,b^{3}+8\,c\,b^{2}
% END_COMPACT Q_V033
\\[1pt]
% BEGIN_COMPACT Q_V223
12\,Q_{V_{112}}={}&-282\,U_{001}+42\,U_{01}\,a+24\,U_{011}+99\,b\,a^{2}-4\,a\,\zeta_2
% END_COMPACT Q_V223
\\[1pt]
% BEGIN_COMPACT Q_V233
6\,Q_{V_{122}}={}&24\,U_{001}-36\,U_{01}\,a+9\,U_{01}\,b+6\,U_{011}-6\,a^{3}+24\,b\,a^{2}-9\,a\,b^{2}+106\,a\,\zeta_2
% END_COMPACT Q_V233
\end{align*}

\noindent\textit{Weight two.}
\begin{align*}
% BEGIN_COMPACT Q_U01
24\,Q_{U_{01}}={}&-36\,U_{01}^{2}-9\,U_{01}\,a^{2}+6\,U_{01}\,a\,b+270\,U_{01}\,a\,c+42\,U_{01}\,a\,f-81\,U_{01}\,b^{2}+66\,U_{01}\,b\,c\\*
&{}-12\,U_{01}\,b\,f-120\,U_{01}\,c^{2}+12\,U_{01}\,c\,f-12\,U_{01}\,e\,f+3\,U_{01}\,f^{2}-90\,U_{01}\,\zeta_2+326\,\zeta_2^{2}\\*
&{}+\zeta_2\,(-850\,a^{2}-114\,a\,b-2\,a\,c+60\,a\,f+360\,b^{2}-24\,b\,c-18\,b\,f+12\,c\,f)\\*
&{}+\zeta_3\,(672\,a-144\,b+12\,f)-70\,a^{4}+30\,b\,a^{3}+154\,c\,a^{3}-46\,f\,a^{3}+84\,a^{2}\,b^{2}-222\,b\,c\,a^{2}\\*
&{}-6\,b\,f\,a^{2}+27\,a^{2}\,c^{2}-48\,c\,f\,a^{2}-30\,a^{2}\,f^{2}+10\,a\,b^{3}+12\,a\,c\,b^{2}+72\,a\,f\,b^{2}-30\,a\,b\,c^{2}+84\,a\,b\,c\,f\\*
&{}+162\,a\,b\,e\,f-48\,a\,b\,f^{2}+60\,a\,c\,e\,f-12\,a\,c\,f^{2}+66\,a\,f\,e^{2}-12\,a\,e\,f^{2}-316\,b^{4}+406\,c\,b^{3}+6\,f\,b^{3}\\*
&{}+54\,b^{2}\,c^{2}-18\,c\,f\,b^{2}-24\,b\,c\,e\,f
% END_COMPACT Q_U01
\\[1pt]
% BEGIN_COMPACT Q_V02
24\,Q_{V_{01}}={}&144\,U_{01}^{2}+U_{01}\,(-36\,V_{01}-96\,a^{2}+84\,a\,b-84\,a\,c+48\,a\,f+540\,b^{2}+672\,b\,c-36\,b\,f-12\,c\,f)\\*
&{}+V_{01}\,(54\,a^{2}+30\,a\,b-12\,a\,c-36\,\zeta_2)-278\,a^{4}+138\,b\,a^{3}+9\,c\,a^{3}-6\,f\,a^{3}-30\,a^{2}\,b^{2}+156\,b\,c\,a^{2}\\*
&{}+123\,b\,f\,a^{2}-6\,a^{2}\,c^{2}-18\,c\,f\,a^{2}-66\,a\,b^{3}-222\,a\,c\,b^{2}-36\,a\,f\,b^{2}+12\,a\,b\,c\,f-18\,a\,b\,f^{2}\\*
&{}+\zeta_2\,(-24\,a\,c+66\,a\,f)
% END_COMPACT Q_V02
\\[1pt]
% BEGIN_COMPACT Q_V03
8\,Q_{V_{02}}={}&76\,U_{01}^{2}+40\,U_{01}\,V_{01}-16\,U_{01}\,V_{02}+144\,U_{01}\,a^{2}+44\,U_{01}\,a\,b-20\,U_{01}\,a\,c\\*
&{}-36\,U_{01}\,a\,f-100\,U_{01}\,b^{2}-4\,U_{01}\,b\,c-8\,U_{01}\,b\,e+8\,U_{01}\,b\,f-4\,U_{01}\,c\,f+64\,U_{01}\,\zeta_2\\*
&{}+V_{01}\,(-20\,a^{2}+10\,a\,b+20\,a\,c-12\,b^{2}-4\,b\,c-6\,b\,f+38\,\zeta_2)+V_{02}\,(12\,a\,b-18\,b^{2}-2\,b\,c)\\*
&{}+\zeta_2\,(244\,a^{2}-212\,a\,b-24\,a\,f+39\,b^{2}+8\,b\,c)+\zeta_3\,(-184\,a-4\,b)-86\,b\,a^{3}-40\,e\,a^{3}-34\,a^{2}\,b^{2}\\*
&{}-66\,b\,c\,a^{2}+6\,b\,e\,a^{2}-14\,b\,f\,a^{2}-42\,c\,e\,a^{2}-8\,e\,f\,a^{2}+65\,a\,b^{3}-122\,a\,c\,b^{2}+8\,a\,e\,b^{2}-36\,a\,f\,b^{2}\\*
&{}+16\,a\,b\,c\,f+12\,b^{4}+6\,c\,b^{3}+4\,e\,b^{3}+3\,f\,b^{3}-4\,c\,f\,b^{2}
% END_COMPACT Q_V03
\\[1pt]
% BEGIN_COMPACT Q_V23
24\,Q_{V_{12}}={}&54\,U_{01}^{2}+U_{01}\,(-48\,a^{2}-306\,a\,b-84\,a\,c-108\,a\,e+96\,a\,f-36\,b^{2}-18\,b\,f+54\,\zeta_2)-264\,V_{02}\,a^{2}\\*
&{}+684\,a\,\zeta_3+\zeta_2\,(862\,a^{2}-108\,a\,b-188\,a\,c-36\,b^{2})+70\,a^{4}+42\,b\,a^{3}-184\,c\,a^{3}-6\,e\,a^{3}-32\,f\,a^{3}\\*
&{}-120\,a^{2}\,b^{2}+42\,b\,c\,a^{2}-198\,b\,e\,a^{2}-48\,b\,f\,a^{2}-21\,a^{2}\,c^{2}+12\,a\,b^{3}+36\,a\,f\,b^{2}
% END_COMPACT Q_V23
\end{align*}

\noindent\textit{Logarithms and periods.}
\begin{align*}
% BEGIN_COMPACT ell_1
12\,\ell_1={}&23\,c
% END_COMPACT ell_1
\qquad % BEGIN_COMPACT ell_2
24\,\ell_2={}139\,b^{2}-296\,b\,c-52\,b\,f-2\,c^{2}-52\,e\,f-323\,\zeta_2
% END_COMPACT ell_2
\\[1pt]
% BEGIN_COMPACT ell_3
48\,\ell_3={}&\zeta_2\,(18\,b+1528\,c+192\,f)+78\,\zeta_3-102\,b^{3}+141\,c\,b^{2}-34\,f\,b^{2}-15\,b\,c^{2}-108\,b\,c\,f-32\,b\,e\,f\\*
&{}-32\,b\,f^{2}+10\,c^{3}-108\,c\,e\,f+8\,f\,e^{2}-32\,e\,f^{2}
% END_COMPACT ell_3
\\[1pt]
% BEGIN_COMPACT ell_4
\ell_4={}&-\tfrac{2503}{80}\,\zeta_2^{2}+\zeta_2\,(\tfrac{19}{8}\,b^{2}-\tfrac{553}{48}\,b\,c-\tfrac{21}{4}\,b\,f-\tfrac{11}{12}\,c^{2}+\tfrac{77}{48}\,c\,e+\tfrac{1}{4}\,c\,f-6\,e\,f+\tfrac{5}{4}\,f^{2})\\*
&{}+\zeta_3\,(-\tfrac{73}{4}\,b+\tfrac{161}{4}\,c-\tfrac{21}{2}\,f)-\tfrac{15}{16}\,b^{4}+\tfrac{11}{8}\,c\,b^{3}+\tfrac{131}{48}\,f\,b^{3}-b^{2}\,c^{2}-\tfrac{7}{8}\,c\,f\,b^{2}+\tfrac{29}{4}\,e\,f\,b^{2}+\tfrac{41}{16}\,b\,f\,c^{2}\\*
&{}-\tfrac{5}{4}\,b\,c\,e\,f+\tfrac{33}{8}\,b\,f\,e^{2}+\tfrac{41}{16}\,e\,f\,c^{2}
% END_COMPACT ell_4
\\[1pt]
% BEGIN_COMPACT ell_5
\ell_5={}&\zeta_2^{2}\,(-\tfrac{2677}{120}\,b+\tfrac{59}{6}\,c-\tfrac{23}{20}\,f)+\zeta_2\,(-\tfrac{32}{9}\,b^{3}+\tfrac{5}{2}\,c\,b^{2}+\tfrac{7}{8}\,f\,b^{2}+\tfrac{23}{6}\,b\,c\,e-\tfrac{1}{2}\,b\,c\,f+\tfrac{9}{4}\,b\,e\,f)\\*
&{}+\zeta_3\,(\tfrac{45}{4}\,b^{2}-\tfrac{253}{12}\,b\,c-b\,f+\tfrac{107}{12}\,c\,e+\tfrac{65}{2}\,c\,f+\tfrac{109}{12}\,\zeta_2)+\tfrac{3}{4}\,\zeta_5+\tfrac{679}{240}\,b^{5}-\tfrac{161}{48}\,c\,b^{4}+\tfrac{1}{4}\,f\,b^{4}-\tfrac{1}{12}\,b^{3}\,c^{2}\\*
&{}+\tfrac{1}{4}\,c\,f\,b^{3}-\tfrac{3}{8}\,b^{3}\,f^{2}+\tfrac{1}{2}\,c\,e\,f\,b^{2}-\tfrac{3}{4}\,e\,b^{2}\,f^{2}
% END_COMPACT ell_5
\\[1pt]
% BEGIN_COMPACT ell_6
840\,\ell_6={}&140\,b\,\zeta_3\,(59\,b^{2}-168\,b\,c+9\,b\,f-6\,c\,f-204\,\zeta_2)-420\,b\,\zeta_5-11474\,\zeta_2^{3}+\zeta_2^{2}\,(1071\,b^{2}+1008\,b\,f)\\*
&{}+\zeta_2\,(-665\,b^{4}+70\,c\,b^{3}-1050\,c\,e\,b^{2})+34650\,\zeta_3^{2}
% END_COMPACT ell_6
\end{align*}

\endgroup
\clearpage
\section{Coproduct spaces and the planar comparison}
\label{app:coproduct}
This appendix gives the symbol-space comparisons summarized in
Section~\ref{sec:comparison}, together with their full-function period
extensions. We use the normalization~\eqref{eq:norm}, the alphabet
ordering~\eqref{eq:alphabet}, and the coproduct convention
$F^{a,b}=(F^b)^a$. Quotients by products are taken over $\mathbb Q$.

\subsection{Products, finite normalization and coproduct ranks}
\label{app:quotient}
For a weight-$k$ symbol $S$, let $\pi_k S$ denote its class modulo
shuffle products of symbols of positive weight; we also write this
class as $[S]$. This projection removes terms that are products of
lower-weight functions. The ordinary symbol already discards positive-weight constants
multiplying lower-weight functions.

Expanding~\eqref{eq:planar-normalizations} to three loops gives
\begin{equation}
 \mathcal E^{(3)}=R^{(3)}+\mathcal E^{(1)}R^{(2)}
 +\frac16(\mathcal E^{(1)})^3-2\zeta_2(\mathcal E^{(1)})^2
 +22\zeta_4\mathcal E^{(1)}.
 \label{eq:planarexponential}
\end{equation}
Consequently, the two planar symbols have the same class in the
weight-six quotient. The SLC class is independent of this common
planar class:
\begin{equation}
 \boxed{[\mathcal S(F)]\notin
 \operatorname{span}_{\mathbb Q}\{[\mathcal S(R^{(3)})]\}.}
 \label{eq:notplanar}
\end{equation}
A short certificate uses the Lyndon Lie brackets associated with the
words $000003$ and $000014$, with letters numbered from zero as
in~\eqref{eq:alphabet}. Their pairings with
$(\mathcal S(\mathcal E^{(3)}),\mathcal S(F))$ are respectively
$(120,0)$ and $(36,-2)$. Lie polynomials annihilate shuffle products.
The determinant $-240$ therefore proves~\eqref{eq:notplanar}, without
requiring a choice of function basis.

The dimensions of the left-factor spaces in~\eqref{eq:Vk} and the
ranks of two projected cobrackets are
\begin{equation}
\begin{array}{c|rrrrr|rr}
 &\multicolumn{5}{c|}{\dim V_k}&
 \operatorname{rank}\delta_{2,4}&\operatorname{rank}\delta_{3,3}\\
 &k=1&2&3&4&5&&\\\hline
 \mathcal E^{(3)}&3&9&12&6&3&5&12\\
 R^{(3)}&3&9&21&9&3&5&12\\
 F&3&9&24&21&5&6&16
\end{array}.
\label{eq:copranks}
\end{equation}
We use
\begin{equation}
 \delta_{p,q}=(\pi_p\otimes\pi_q)
       (\Delta_{p,q}-\tau\Delta_{q,p}),
 \label{eq:appendix-cobracket}
\end{equation}
where $\tau$ exchanges the tensor factors~\cite{Duhr,GPSV}.
The last two columns of~\eqref{eq:copranks} are matrix ranks of
individual cobracket tensors. The dimensions of the images
$\delta_{2,2}V_4$ in~\eqref{eq:V4cobrackets}, in contrast, count
independent images of an entire left-factor space.

\subsection{Weight-three symbols and their period extensions}
\label{sec:planarperiods}\label{sec:sharedspaces}
The unprojected weight-three spaces differ by three explicit
products:
\begin{equation}
 \begin{split}
 V_3(\mathcal S(F))={}&V_3(\mathcal S(R^{(3)}))\oplus\\
 &\operatorname{span}_{\mathbb Q}\!\left\{
 \mathcal S\bigl(\log u_i\,\operatorname{Li}_2(1-u_i)\bigr):
 u_i=u,v,w\right\}.
 \end{split}
 \label{eq:V3products}
\end{equation}
The three displayed product directions are independent modulo the
planar remainder space. Adjoining them increases its dimension from 21
to 24, so the sum is direct.
After applying $\pi_3$, all three spaces in~\eqref{eq:V3modproducts}
coincide and have dimension eight.

This SLC space also has a description independent of the three-loop
coefficients. Let $\mathcal M_3^{\rm symb}$ be the 27-dimensional
physical form-factor symbol space of Ref.~\cite{DMW}, without imposing
permutation invariance. Define $D_i(X)=[i\otimes i\otimes i]X$ for
$i=u,v,w$. Then
\begin{equation}
 V_3(\mathcal S(F))=\ker(D_u,D_v,D_w)
 \subset\mathcal M_3^{\rm symb},\qquad \dim V_3=24.
 \label{eq:leftkernel}
\end{equation}
The three maps $D_i$ are independent on $\mathcal M_3^{\rm symb}$.
All SLC weight-three left factors lie in their common kernel and
span 24 dimensions, establishing the equality.

The two-loop planar remainder supplies elements of the same kernel:
$\mathcal S((R_{\rm LC}^{(2)})^j)\in V_3(\mathcal S(F))$.
The BDS-like normalization supplies a complementary space instead,
since
\begin{equation}
 D_i\!\left(\mathcal S((\mathcal E^{(2)})^j)\right)
       =12\delta_{ij},\qquad i,j=u,v,w.
 \label{eq:two-loop-complement}
\end{equation}
Thus any $X\in\mathcal M_3^{\rm symb}$ can be projected onto the SLC
space by subtracting
$\tfrac1{12}\sum_i D_i(X)\mathcal S((\mathcal E^{(2)})^i)$.
This comparison depends on the finite normalization: the lower-loop
products relating $R^{(2)}$ and $\mathcal E^{(2)}$ restore the three
pure logarithmic directions detected by $D_i$.

At full-function level the weight-three left factors span a different
27-dimensional space, denoted $V_3^{\rm fun}(F)$. Its symbol image is
the 24-dimensional kernel in~\eqref{eq:leftkernel}. The kernel of the
symbol map on this function space is
\begin{equation}
 \boxed{\ker\mathcal S\big|_{V_3^{\rm fun}(F)}
 =\operatorname{span}_{\mathbb Q}
 \left\{\zeta_2\log u_i-\frac52\zeta_3:
 u_i=u,v,w\right\}.}
 \label{eq:periodkernel}
\end{equation}
The function identities in this appendix are understood on the
Euclidean branch.
The fixed $\zeta_3$ shifts tie the constant to the three logarithmic
directions. In particular, a pure $\zeta_3$ is not contained in this
space: a constant linear combination of $\log u,\log v,\log w$ can
vanish for all $u+v+w=1$ only if all coefficients vanish.
Adjoining $\zeta_3$ increases the dimension to 28.

With the normalization~\eqref{eq:planar-ambiguity}, the precise
embedding of the two-loop functions is
\begin{equation}
 (R_{\rm LC}^{(2)})^i-6\zeta_3\in V_3^{\rm fun}(F),\qquad i=u,v,w.
 \label{eq:R2embedding}
\end{equation}
Their pairwise differences lie in $V_3^{\rm fun}(F)$ without a shift;
the sum requires the constant $18\zeta_3$. These function-space
relations are distinct from both the symbol quotient and the
weight-six ambiguity $K_6$ of Section~\ref{sec:uplift}.

\subsection{The additional weight-four directions}
\label{app:weightfour}
Let $\mathcal M_4^{\rm symb}$ contain all integrable weight-four
symbols in~\eqref{eq:alphabet}, with first entries in $\{u,v,w\}$,
no adjacent distinct complementary letters, and the recursive branch
conditions~\eqref{eq:pseudothresh}. No permutation symmetry is imposed. This space has dimension 81, and
\begin{equation}
 \delta_{2,2}\mathcal M_4^{\rm symb}
 =\delta_{2,2}V_4(\mathcal S(F)),\qquad \dim=9.
 \label{eq:M4cobracket}
\end{equation}
The 21-dimensional space $V_4(\mathcal S(F))$ and the 81-dimensional
space $\mathcal M_4^{\rm symb}$ thus have the same nine-dimensional
nonclassical image. Classical and product components are removed by
$\delta_{2,2}$.

The four directions beyond the planar image have a compact basis.
For $(i,j,k)=(u,v,w),(v,w,u),(w,u,v)$, set
\begin{equation}
 d_i=\pi_2\mathcal S\bigl(\operatorname{Li}_2(1-i)\bigr),\qquad
 e_i=\pi_2\mathcal S\bigl(\operatorname{Li}_2(-j/k)\bigr),\qquad
 \Xi_i=d_i\wedge e_i,
 \label{eq:dilog-classes}
\end{equation}
where $\wedge$ is the antisymmetric tensor product, and define
\begin{equation}
 \Upsilon=2(d_u\wedge d_v+d_v\wedge d_w+d_w\wedge d_u)
 +(d_u+d_v+d_w)\wedge(e_u+e_v+e_w).
 \label{eq:dilog-sign}
\end{equation}
The four classes $\Xi_u,\Xi_v,\Xi_w,\Upsilon$ are independent and
span a complement to the common five-dimensional planar image in
\eqref{eq:V4cobrackets}.

To relate them to final entries, define
\begin{equation}
 J_i=\mathcal S(F)^i+\mathcal S(F)^{\bar i},\qquad i=u,v,w.
 \label{eq:J-deviations}
\end{equation}
We also set $K_i=(J_i)^{\bar i}$. The sum of the $J_i$ vanishes by
\eqref{eq:weaklast}; the corresponding function-level sum is $C_5$
in~\eqref{eq:lastsource}. With the sum over cyclic pairs
$(i,j)=(u,v),(v,w),(w,u)$, introduce
\begin{align}
 \Omega_0&=\sum_{(i,j)\,\mathrm{cyc}}
       \bigl[(J_i)^j-(J_j)^i\bigr],\nonumber\\
 \Omega_1&=\sum_{(i,j)\,\mathrm{cyc}}
       \bigl[(J_i)^{\bar j}-(J_j)^{\bar i}\bigr].
 \label{eq:Omega01}
\end{align}
The images of $\Omega_0,\Omega_1,K_u-K_v,K_v-K_w$ give the same
four-dimensional complement. The two bases are related by
\begin{align}
 \delta_{2,2}(\Omega_1-\tfrac12\Omega_0)
     &=9(\Xi_u+\Xi_v+\Xi_w),&
 \delta_{2,2}(\Omega_1-2\Omega_0)&=6\Upsilon,\nonumber\\
 \delta_{2,2}(K_u-K_v)&=2\Xi_w-\Xi_u-\Xi_v,&
 \delta_{2,2}(K_v-K_w)&=2\Xi_u-\Xi_v-\Xi_w.
 \label{eq:dilog-sources}
\end{align}
These equations express the additional nonclassical directions in
terms of coproduct components of the single-final-entry deviations.

The dilogarithm inversion identity makes $e_i$ transform as
$-e_{p(i)}$ under an odd permutation $p$, modulo products and periods.
The three $\Xi_i$ therefore carry a sign representation and a
standard representation of $S_3$; $\Upsilon$ carries a second sign
representation. Equivalently, $\Omega_0$ and $\Omega_1$ change sign
under odd permutations, while the two $K_i$ differences form the
standard representation. The complement is
$2\,\mathrm{sgn}\oplus\mathrm{std}$.

The indecomposable classes of both planar weight-six symbols also
admit permutation-sum representations in the five-letter
alphabet~\eqref{eq:seed-alphabet}. For the SLC correction, the same
alphabet organizes the full function, including products and periods,
as in~\eqref{eq:master-explicit}.

\section{Last-two adjacency and higher final-entry relations}
\label{app:finalentries}
This appendix collects the full-function identities behind the
last-two structure, their use in the bootstrap and the finite-subtraction
obstruction. All functions are evaluated at $Q^2=1$ in the Euclidean
region unless stated otherwise.

\subsection{Adjacency defects and integrability}
The numbers of words violating complementary-letter adjacency in the
five successive pairs of symbol slots are
\begin{equation}
 \mathcal E^{(3)}:(0,0,0,0,0),\qquad
 R^{(3)}:(0,0,72,216,312),\qquad F:(0,0,0,0,468).
 \label{eq:countsadj}
\end{equation}
The comparison distinguishes the planar BDS-like quantity from the
planar remainder. For $F$, only the final pair violates adjacency.
The six ordered complementary pairs reduce to three weight-four
functions,
\begin{equation}
 A_{uv}=F^{\bar u,\bar v},\qquad
 A_{uw}=F^{\bar u,\bar w},\qquad A_{vw}=F^{\bar v,\bar w}.
 \label{eq:defects}
\end{equation}
These functions are independent over $\mathbb Q$ and have 78 symbol
words each.

At the symbol level, the independent
$d\log\bar u\wedge d\log\bar v$ component of integrability equates
the coefficients with these two final entries exchanged. Weak
adjacency then excludes every complementary final entry from
$A_{uv}$. Indeed, a coefficient
$F^{\bar c,\bar u,\bar v}$ vanishes directly if $c\ne u$.
For $c=u$, exchanging the last two entries gives
$F^{\bar u,\bar v,\bar u}$, which vanishes by adjacency in the
preceding pair. The full-function coproducts satisfy the same
identities:
\begin{equation}
 A_{uv}=A_{vu},\qquad
 A_{uv}^{\bar u}=A_{uv}^{\bar v}=A_{uv}^{\bar w}=0.
 \label{eq:defectsimple}
\end{equation}
The integrability argument establishes the symbol identities; their
full-function counterparts follow from the reconstructed expression.

\subsection{Low-weight sources and an alternative reconstruction}
\label{app:sources}
The remaining coproducts obey the logarithmic and box relations
\begin{align}
 A_{uv}^{u}+A_{uv}^{w}&=\frac12\log^2v\log\frac uw,\nonumber\\
 A_{uv}^{v}+A_{uv}^{w}&=\frac12\log^2u\log\frac vw,\nonumber\\
 A_{uv}^{w,w}&=b_\Box(u,v)
 =\zeta_2-\operatorname{Li}_2(1-u)-\operatorname{Li}_2(1-v)-\log u\log v.
 \label{eq:defectsources}
\end{align}
The box function is equivalently
\begin{equation}
 b_\Box(u,v)=\operatorname{Li}_2(1-1/u)
       +\operatorname{Li}_2(1-1/v)
       +\frac12\log^2\frac uv+\zeta_2.
 \label{eq:box-finite-convention}
\end{equation}
It is the finite one-mass-box function with the infrared powers
kept separate~\cite{BDDK}. The equality of the two expressions uses
$\operatorname{Li}_2(1-1/u)=-\operatorname{Li}_2(1-u)
-\tfrac12\log^2u$ on the stated branch.

Equation~\eqref{eq:defectsimple} gives
$dA_{uv}=A_{uv}^{u}d\log u+A_{uv}^{v}d\log v
+A_{uv}^{w}d\log w$. Since $w=1-u-v$,
\begin{equation}
 \partial_u A_{uv}=\frac{A_{uv}^u}{u}-\frac{A_{uv}^w}{w},
 \qquad
 \partial_v A_{uv}=\frac{A_{uv}^v}{v}-\frac{A_{uv}^w}{w}.
 \label{eq:defect-derivative}
\end{equation}
Acting on $A_{uv}$, the operator
$u(1-u)\partial_u-v(1-v)\partial_v$ gives
$(1-u)(A_{uv}^u+A_{uv}^w)-(1-v)(A_{uv}^v+A_{uv}^w)$.
Substituting the logarithmic sources gives
\begin{equation}
 \begin{split}
 &[u(1-u)\partial_u-v(1-v)\partial_v]A_{uv}\\
 &\quad=\frac{1-u}{2}\log^2v\log\frac uw
       -\frac{1-v}{2}\log^2u\log\frac vw.
 \end{split}
 \label{eq:AuvPDE}
\end{equation}
Its characteristic invariant is $uv/w$, so its homogeneous solutions
are arbitrary functions of that ratio. The remaining coproduct and
boundary conditions restrict this freedom. Equation~\eqref{eq:AuvPDE} is a first-order partial differential
equation for $A_{uv}(u,v)$, equivalent to the stated combination of
coproduct identities.

For the full symbol bootstrap, define
\begin{equation}
 \mathcal L_{uv}[G]=G^{u,\bar u,\bar v}+G^{w,\bar u,\bar v},\qquad
 X_{uv}=\frac12\log^2v\log\frac uw.
 \label{eq:logsource-operator}
\end{equation}
The equation $\mathcal L_{uv}[F]=X_{uv}$ and its permutations reduce
the 45 parameters after collinear matching to 36. On that candidate
space their homogeneous kernel is identical to the kernel obtained
by fixing all three defect symbols in~\eqref{eq:defects}. Adding
$T_1,T_2,T_3$ gives the exact sequence
\begin{equation}
 45\ \xrightarrow{\,\mathcal L_{ij}\,}\ 36
 \ \xrightarrow{\,T_1\,}\ 18
 \ \xrightarrow{\,T_2\,}\ 9
 \ \xrightarrow{\,T_3\,}\ 0.
 \label{eq:source-bootstrap-sequence}
\end{equation}
All permutation images are included. The final zero denotes a
unique solution of the inhomogeneous system; $T_4$ can replace $T_3$
in the last step.

At the function level, the same source detects the last ambiguity. For
the particular solution and kernel used in
Section~\ref{sec:uplift},
\begin{equation}
 \mathcal L_{uv}[K_6]=-\frac{\zeta_2}{2}\log v,\qquad
 \mathcal L_{uv}[F_{\rm part}]-X_{uv}=-\frac{\zeta_2}{10}\log v.
 \label{eq:source-uplift}
\end{equation}
Thus $F_{\rm part}+cK_6$ satisfies the source equation precisely for
$c=-1/5$, the value selected by $T_1[F]=0$. The source relation involves
cancellations between the weight-six GPL core and the product and
period terms of the seed. When $A_{uv}$ is considered independently,
its source equations still allow
$c_3\zeta_3\log(uv/w)+c_4\zeta_4$; the full form-factor reconstruction
fixes these constants.

\Needspace{24\baselineskip}
\subsection{A classical expression for the defect}
\label{app:classical}
On the Euclidean branch, $A_{uv}$ can be expressed entirely in
classical polylogarithms. Set
\begin{equation}
 a=\log u,\quad b=\log v,\quad c=\log w,\quad
 t=a+b-c,\qquad z=-\frac{uv}{w},
 \label{eq:classical-defect-vars}
\end{equation}
and define
\begin{align}
 \chi_4(u,v)={}&-3\bigl[\operatorname{Li}_4(u)
                  +\operatorname{Li}_4(-u/w)\bigr]
 +(3a-t)\operatorname{Li}_3(u)\nonumber\\
 &+(a-2b-c)\operatorname{Li}_3(-u/w)
 +a\left(b-c-\frac a2\right)
       \bigl[\operatorname{Li}_2(u)+\operatorname{Li}_2(-v/w)\bigr],
 \label{eq:chi4}\\
 \mathcal P_4(a,b,c)={}&\frac14a^2b^2-\frac14(a-b)^2c^2
 -\frac16(a+b)c^3+\frac1{24}c^4
 -\frac{\zeta_2}{2}t^2+\zeta_3t-\frac94\zeta_4.
 \label{eq:defect-polynomial}
\end{align}
Then
\begin{equation}
 \boxed{A_{uv}=\chi_4(u,v)+\chi_4(v,u)
       +3\operatorname{Li}_4(z)-t\operatorname{Li}_3(z)
       +\mathcal P_4(a,b,c).}
 \label{eq:classical-defect}
\end{equation}
The arguments $u,v,-u/w,-v/w,-uv/w$ and their complements involve
only $\{u,v,w,1-u,1-v\}$. For example,
\begin{equation}
 1+\frac uw=\frac{1-v}{w},\qquad
 1+\frac vw=\frac{1-u}{w},\qquad
 1+\frac{uv}{w}=\frac{(1-u)(1-v)}{w}.
 \label{eq:defect-alphabet-closure}
\end{equation}
These factorizations make the five-letter closure explicit.
Equation~\eqref{eq:classical-defect} includes the period terms required
for equality of functions. Its $\delta_{2,2}$ vanishes, consistently
with the weight-four classicality criterion~\cite{GSVV,DGR}.
Classicality does not make the function decomposable: its
$\operatorname{Li}_4$ part is nonzero modulo lower-weight products,
as the certificate below shows.

\subsection{Finite subtraction and the product obstruction}
\label{sec:subtraction}
A color-dependent finite change of the infrared subtraction factor
acts as $F\to F-C_6^{\rm fin}$. Multiplication by a common color-independent finite factor with
unit tree-level term leaves the first SLC coefficient unchanged, as
does replacing $H$ by $\log H$.

Consider a constant-coefficient linear combination $C_6^{\rm fin}$
of products of lower-weight pure functions, each factor having a
physical first entry. Define
\begin{equation}
 \Omega_{uv}(X)=\pi_4\,\mathcal S(X^{\bar u,\bar v}),\qquad
 \Omega_{uv}(C_6^{\rm fin})=0.
 \label{eq:Omega}
\end{equation}
To prove the vanishing, project the multiplicative coproduct onto its
$(4,2)$ component and select $\bar u\otimes\bar v$ on the right.
If at least two positive-weight factors remain on the left, their
shuffle product is annihilated by $\pi_4$. If only one positive-weight
factor remains on the left, at least one other nonconstant factor must
be placed entirely on the right. Such a factor begins with $u,v$ or
$w$, and cannot contribute to a word consisting of two complementary
letters. Positive-weight constant factors have zero ordinary symbol.
These cases exhaust the product coproduct and establish
$\Omega_{uv}(C_6^{\rm fin})=0$.

For the actual defect,
\begin{equation}
 \left\langle [u,[u,[v,w]]],\mathcal S(A_{uv})\right\rangle=-3.
 \label{eq:Liecertificate}
\end{equation}
The bracket here is a formal Lie polynomial in the tensor letters;
it annihilates every shuffle product. Hence
\begin{equation}
 F\ne G_6+\text{physical-first-entry lower-weight products},
 \qquad G_6\text{ strictly adjacent}.
 \label{eq:nogofinite}
\end{equation}
The obstruction is an indecomposable weight-four component of the
last-two defect. By contrast, the two-loop planar normalization change
cancels a product:
\begin{equation}
 (R^{(2)})^{\bar u,\bar v}=-4\log u\log v,\qquad
 \left[\tfrac12(\mathcal E^{(1)})^2\right]^{\bar u,\bar v}
 =4\log u\log v.
 \label{eq:planar-product-cancellation}
\end{equation}
A finite subtraction restoring SLC adjacency must therefore lie
outside the stated class of lower-weight products.

There is a separate boundary obstruction to restoring all three
planar single-final-entry equations. Suppose $P^i+P^{\bar i}=0$
for $i=u,v,w$. Then $dP=\sum_iP^i\,d\log(i/\bar i)$.
On the collinear boundary
$(u,v,w)=(\delta,z,1-z-\delta)$, the last entries reduce to
$d\log\delta$ and $d\log(z/(1-z))$. The leading-power boundary $B$ of $P$
must consequently satisfy
\begin{equation}
 B^z+B^{1-z}=0.
 \label{eq:planar-boundary-obstruction}
\end{equation}
The SLC collinear symbol does not satisfy this relation. A finite
normalization restoring the three planar final-entry conditions must
therefore change the collinear boundary as well as the bulk function.

\subsection{Four-final relations}
\label{sec:fourfinal}
With integrability, the weak single-final-entry equation and the
chosen adjacency restriction imposed, no further constant-coefficient
homogeneous relations arise at depths one or two. Additional relations
first appear at depth three. At depth four we find the following
four two-term families:
\begin{align}
 Q_1[F]&=F^{u,u,v,v}+F^{v,u,v,v},&
 Q_2[F]&=F^{u,v,\bar v,w}-F^{v,\bar w,v,\bar u},\nonumber\\
 Q_3[F]&=F^{u,v,\bar v,w}-F^{\bar w,u,u,\bar v},&
 Q_4[F]&=F^{u,w,\bar w,v}-F^{w,\bar v,v,u}.
 \label{eq:Q}
\end{align}
The equations $Q_j[F]=0$ and all their permutation images hold as
identities between full weight-two functions, with no additional
$\zeta_2$ terms. Neither planar three-loop
normalization satisfies these four families.

On the 45-parameter space left after collinear matching, the $Q_j$
and their permutations leave 18 parameters. Adding $T_1$ fixes the
symbol. The same $Q_j$, however, satisfy
\begin{equation}
 \boxed{Q_j[K_6]=0,\qquad j=1,\ldots,4.}
 \label{eq:Qblind}
\end{equation}
Thus their full-function validity does not remove the
$\zeta_2R_{\rm LC}^{(2)}/8$ ambiguity. The $T_1$ relation or the
logarithmic source in~\eqref{eq:source-uplift} supplies that last
constraint. At total weight six the four-final components have
weight two, whereas at total weight eight they have weight four.
The $Q_j$ are therefore used here as three-loop identities, alongside
the higher-final-entry structures known in planar
bootstraps~\cite{EightLoops}.

\clearpage
\begingroup\fontsize{10}{11.3}\selectfont

\endgroup

\begin{thebibliography}{99}
\setlength{\itemsep}{0pt}\setlength{\parskip}{0pt}
\bibitem{GSVV}
A.~B.~Goncharov, M.~Spradlin, C.~Vergu and A.~Volovich,
\emph{Classical Polylogarithms for Amplitudes and Wilson Loops},
Phys.\ Rev.\ Lett.\ \textbf{105} (2010) 151605,
\href{https://arxiv.org/abs/1006.5703}{arXiv:1006.5703}.

\bibitem{Form2}
A.~Brandhuber, G.~Travaglini and G.~Yang,
\emph{Analytic two-loop form factors in $\mathcal N=4$ SYM},
JHEP \textbf{05} (2012) 082,
\href{https://arxiv.org/abs/1201.4170}{arXiv:1201.4170}.

\bibitem{DGR}
C.~Duhr, H.~Gangl and J.~R.~Rhodes,
\emph{From polygons and symbols to polylogarithmic functions},
JHEP \textbf{10} (2012) 075,
\href{https://arxiv.org/abs/1110.0458}{arXiv:1110.0458}.

\bibitem{DMW}
L.~J.~Dixon, A.~J.~McLeod and M.~Wilhelm,
\emph{A Three-Point Form Factor Through Five Loops},
JHEP \textbf{04} (2021) 147,
\href{https://arxiv.org/abs/2012.12286}{arXiv:2012.12286}.

\bibitem{EightLoops}
L.~J.~Dixon, O.~G\"urdo\u{g}an, A.~J.~McLeod and M.~Wilhelm,
\emph{Bootstrapping a Stress-Tensor Form Factor through Eight Loops},
JHEP \textbf{07} (2022) 153,
\href{https://arxiv.org/abs/2204.11901}{arXiv:2204.11901}.

\bibitem{FullColor5}
S.~Abreu, L.~J.~Dixon, E.~Herrmann, B.~Page and M.~Zeng,
\emph{The two-loop five-point amplitude in $\mathcal N=4$ super-Yang--Mills theory},
Phys.\ Rev.\ Lett.\ \textbf{122} (2019) 121603,
\href{https://arxiv.org/abs/1812.08941}{arXiv:1812.08941}.

\bibitem{GMR}
X.~Guan, B.~Mistlberger and M.~Ruf,
\emph{Subleading Color Corrections at Three Loops to the
$\operatorname{tr}(\phi^2)$ Three-Point Form Factor in $\mathcal N=4$
Super Yang--Mills Theory},
\href{https://arxiv.org/abs/2511.13845}{arXiv:2511.13845}.

\bibitem{GPSV}
J.~Golden, M.~F.~Paulos, M.~Spradlin and A.~Volovich,
\emph{Cluster Polylogarithms for Scattering Amplitudes},
J.\ Phys.\ A \textbf{47} (2014) 474005,
\href{https://arxiv.org/abs/1401.6446}{arXiv:1401.6446}.

\bibitem{PSSV}
D.~E.~Parker, A.~Scherlis, M.~Spradlin and A.~Volovich,
\emph{Hedgehog Bases for $A_n$ Cluster Polylogarithms and An Application to
Six-Point Amplitudes}, JHEP \textbf{11} (2015) 136,
\href{https://arxiv.org/abs/1507.01950}{arXiv:1507.01950}.

\bibitem{FF4SLC}
S.~He and X.~Jiang,
\emph{A full-color four-point form factor},
to appear.

\bibitem{Higgs2}
T.~Gehrmann, M.~Jaquier, E.~W.~N.~Glover and A.~Koukoutsakis,
\emph{Two-Loop QCD Corrections to the Helicity Amplitudes for $H\to3$ partons},
JHEP \textbf{02} (2012) 056,
\href{https://arxiv.org/abs/1112.3554}{arXiv:1112.3554}.

\bibitem{HiggsSYM}
A.~Brandhuber, M.~Kostaci\'nska, B.~Penante and G.~Travaglini,
\emph{Higgs amplitudes from $\mathcal N=4$ super Yang--Mills theory},
Phys.\ Rev.\ Lett.\ \textbf{119} (2017) 161601,
\href{https://arxiv.org/abs/1707.09897}{arXiv:1707.09897}.

\bibitem{Higgs3}
X.~Guan, B.~Mistlberger and M.~S.~Ruf,
\emph{Three-Loop QCD Corrections to Scattering Amplitudes of a Higgs Boson
and Three Partons},
\href{https://arxiv.org/abs/2607.12033}{arXiv:2607.12033}.

\bibitem{Form1}
A.~Brandhuber, B.~Spence, G.~Travaglini and G.~Yang,
\emph{Form Factors in $\mathcal N=4$ Super Yang--Mills and Periodic Wilson Loops},
JHEP \textbf{01} (2011) 134,
\href{https://arxiv.org/abs/1011.1899}{arXiv:1011.1899}.

\bibitem{IRfactor}
E.~Gardi and L.~Magnea,
\emph{Factorization constraints for soft anomalous dimensions in QCD
scattering amplitudes}, JHEP \textbf{03} (2009) 079,
\href{https://arxiv.org/abs/0901.1091}{arXiv:0901.1091}.

\bibitem{ADG}
\O.~Almelid, C.~Duhr and E.~Gardi,
\emph{Three-loop corrections to the soft anomalous dimension in multi-leg scattering},
Phys.\ Rev.\ Lett.\ \textbf{117} (2016) 172002,
\href{https://arxiv.org/abs/1507.00047}{arXiv:1507.00047}.

\bibitem{HighPrecision}
X.~Guan, G.~Lin, X.~Liu, Y.-Q.~Ma and G.~Yang,
\emph{A high-precision result for a full-color three-loop three-point form factor
in $\mathcal N=4$ SYM}, JHEP \textbf{02} (2024) 201,
\href{https://arxiv.org/abs/2309.04395}{arXiv:2309.04395}.

\bibitem{BDS}
Z.~Bern, L.~J.~Dixon and V.~A.~Smirnov,
\emph{Iteration of Planar Amplitudes in Maximally Supersymmetric Yang--Mills
Theory at Three Loops and Beyond},
Phys.\ Rev.\ D \textbf{72} (2005) 085001,
\href{https://arxiv.org/abs/hep-th/0505205}{arXiv:hep-th/0505205}.

\bibitem{Duhr}
C.~Duhr, \emph{Hopf algebras, coproducts and symbols: an application to
Higgs boson amplitudes}, JHEP \textbf{08} (2012) 043,
\href{https://arxiv.org/abs/1203.0454}{arXiv:1203.0454}.

\bibitem{Splitting}
X.~Guan, F.~Herzog, Y.~Ma, B.~Mistlberger and A.~Suresh,
\emph{Splitting amplitudes at N$^3$LO in QCD},
\href{https://arxiv.org/abs/2408.03019}{arXiv:2408.03019}.

\bibitem{Brown}
F.~C.~S.~Brown,
\emph{Multiple zeta values and periods of moduli spaces $\mathfrak M_{0,n}$},
\href{https://arxiv.org/abs/math/0606419}{arXiv:math/0606419}.

\bibitem{Soft3}
W.~Chen, M.-x.~Luo, T.-Z.~Yang and H.~X.~Zhu,
\emph{Soft Theorem to Three Loops in QCD and $\mathcal N=4$ Super Yang--Mills
Theory}, JHEP \textbf{01} (2024) 131,
\href{https://arxiv.org/abs/2309.03832}{arXiv:2309.03832}.

\bibitem{BDDK}
Z.~Bern, L.~J.~Dixon, D.~C.~Dunbar and D.~A.~Kosower,
\emph{One-Loop n-Point Gauge Theory Amplitudes, Unitarity and Collinear Limits},
Nucl.\ Phys.\ B \textbf{425} (1994) 217--260,
\href{https://arxiv.org/abs/hep-ph/9403226}{arXiv:hep-ph/9403226}.

\bibitem{LYZ}
G.~Lin, G.~Yang and S.~Zhang,
\emph{Color-Kinematics Duality and Dual Conformal Symmetry for A Four-loop
Form Factor in N=4 SYM},
\href{https://arxiv.org/abs/2112.09123}{arXiv:2112.09123}.

\bibitem{IR}
C.~Duhr, E.~Gardi, S.~Jaskiewicz, J.~L\"ubken and L.~Vernazza,
\emph{Infrared singularities and the collinear limits of multi-leg
scattering amplitudes},
\href{https://arxiv.org/abs/2507.21854}{arXiv:2507.21854}.

\bibitem{Sudakov4}
R.~N.~Lee, A.~von Manteuffel, R.~M.~Schabinger, A.~V.~Smirnov,
V.~A.~Smirnov and M.~Steinhauser,
\emph{The Four-Loop $\mathcal N=4$ SYM Sudakov Form Factor},
JHEP \textbf{01} (2022) 091,
\href{https://arxiv.org/abs/2110.13166}{arXiv:2110.13166}.

\bibitem{FF4MHV}
L.~J.~Dixon, \"O.~G\"urdo\u{g}an, Y.-T.~Liu, A.~J.~McLeod and M.~Wilhelm,
\emph{Antipodal Self-Duality for a Four-Particle Form Factor},
Phys.\ Rev.\ Lett.\ \textbf{130} (2023) 111601,
\href{https://arxiv.org/abs/2212.02410}{arXiv:2212.02410}.

\bibitem{FF4Function}
L.~J.~Dixon and S.~Xin,
\emph{A Two-Loop Four-Point Form Factor at Function Level},
JHEP \textbf{01} (2025) 012; erratum JHEP \textbf{03} (2026) 159,
\href{https://arxiv.org/abs/2411.01571}{arXiv:2411.01571}.

\bibitem{FF4Analytic}
Y.~Guo, L.~Wang, G.~Yang and Y.~Yin,
\emph{Analytic two-loop four-point form factor of the stress-tensor
supermultiplet in $\mathcal N=4$ SYM},
JHEP \textbf{02} (2025) 002,
\href{https://arxiv.org/abs/2409.12445}{arXiv:2409.12445}.

\bibitem{FF4NMHV}
S.~He, J.~Liu and Q.~Yang,
\emph{Bootstrapping the Four-Point NMHV Stress-Tensor Form Factor},
\href{https://arxiv.org/abs/2605.28955}{arXiv:2605.28955}.

\bibitem{OneMass2}
S.~Abreu, D.~Chicherin, H.~Ita, B.~Page, V.~Sotnikov,
W.~Tschernow and S.~Zoia,
\emph{All Two-Loop Feynman Integrals for Five-Point One-Mass Scattering},
Phys.\ Rev.\ Lett.\ \textbf{132} (2024) 141601,
\href{https://arxiv.org/abs/2306.15431}{arXiv:2306.15431}.

\bibitem{Antipodal}
L.~J.~Dixon, O.~G\"urdo\u{g}an, A.~J.~McLeod and M.~Wilhelm,
\emph{Folding Amplitudes into Form Factors: An Antipodal Duality},
Phys.\ Rev.\ Lett.\ \textbf{128} (2022) 111602,
\href{https://arxiv.org/abs/2112.06243}{arXiv:2112.06243}.

\end{thebibliography}
\end{document}